\documentclass[aps,floatfix,preprint,preprintnumbers,nofootinbib,superscriptaddress,natbib]{revtex4}

\usepackage{graphicx,float}
\usepackage{epsfig}		
\usepackage{dcolumn}
\usepackage{bm}
\usepackage{subfigure}

\usepackage[all]{xy}
\usepackage{amsmath,upgreek}
\usepackage{amssymb}

\usepackage{pdfpages}
\usepackage{color,soul}
\usepackage{graphicx,epstopdf}
\usepackage[colorlinks=true,
 linkcolor=red,
 urlcolor=purple,
 citecolor=blue,hyperindex]{hyperref}
\def\0{\mbox{\tiny $0$}}
\def\1{\mbox{\tiny $1$}}
\def\2{\mbox{\tiny $2$}}
\def\3{\mbox{\tiny $3$}}
\def\4{\mbox{\tiny $4$}}
\def\5{\mbox{\tiny $5$}}
\def\6{\mbox{\tiny $6$}}
\def\7{\mbox{\tiny $7$}}
\def\8{\mbox{\tiny $8$}}
\def\9{\mbox{\tiny $9$}}

\def\f14{\mbox{\tiny $\frac{1}{4}$}}

\begin{document}

\title{Toda-like Hamiltonian as a probe for quantized prey-predator dynamics}


\renewcommand{\baselinestretch}{1.2}
\author{Alex E. Bernardini}
\email{alexeb@ufscar.br}
\altaffiliation[On leave of absence from]{~Departamento de F\'{\i}sica, Universidade Federal de S\~ao Carlos, PO Box 676, 13565-905, S\~ao Carlos, SP, Brasil.}
\author{Orfeu Bertolami}
\email{orfeu.bertolami@fc.up.pt}
\altaffiliation[Also at~]{Centro de F\'isica do Porto, Rua do Campo Alegre 687, 4169-007, Porto, Portugal.} 
\affiliation{Departamento de F\'isica e Astronomia, Faculdade de Ci\^{e}ncias da Universidade do Porto, Rua do Campo Alegre 687, 4169-007, Porto, Portugal.}
\date{\today}

\begin{abstract}
Phase-space features of a reduced version of the Toda-like Hamiltonian, $\mathcal{H}(x,\,k)$, written in a form constrained by the condition $\partial^2 \mathcal{H} / \partial x \partial k = 0$, with $x$ and $k$ as canonically conjugate variables, are analyzed in terms of Wigner currents. 
For Wigner currents convoluted with either thermodynamic or Gaussian ensembles, the underlying Hamiltonian dynamics admits analytic corrections due to quantum distortions over the classical phase-space pattern, computed and interpreted through quantifiers of quantumness and stationarity.
Notably, while emulating the Lotka-Volterra (LV) dynamics that describe ecological competition systems, the Toda-like classical dynamics allows for analytical solutions with computable periods corresponding to closed phase-space orbits of isotropic prey-predator population distributions. 
The essential conditions for understanding how classical and quantum evolution can coexist are provided at different scales of quantumness, driven by the associated convoluting ensemble parameter.
In the case of Gaussian statistical ensembles, the exact profile of the quantum distortions over classical prey-predator phase-space trajectories is obtained non-perturbatively. 
Our results indicate that, besides the classical stability admitted by LV models, the Toda-like patterns also exhibit quantum stability. Therefore, this can be regarded as the first step as a predictive theoretical framework towards more robust descriptions of quantum patterns in competitive microscopic biosystems.
\end{abstract}

\keywords{Phase Space Quantum Mechanics - Wigner Formalism - Prey Predator Dynamics - Quantumness - Toda}

\date{\today}
\maketitle

\section{Introduction}

The interpretation of microscopic ecological system instabilities in terms of the Weyl-Wigner (WW) quantum mechanics (QM) has been investigated through various platforms~\cite{NossoPaper,NossoPaper19,Novo2022,Novo2022B}.
It has been correlated with a wide range of appealing effective models which, for instance, explain the stability criteria for microbiological complex systems~\cite{Nature01, Nature02}, drive the onset of phase transitions in finite microbiological ensembles \cite {PRE-LV3}, as well as introduce more grounded criteria for macroscopic extinction divergence effects~\cite{SciRep02,PRE-LV2}. 

In some sense, the related Wigner flow analysis can also map the Fokker-Planck equation framework, through which distortion effects around (classical) stability and equilibrium conditions are achieved through a mean-field approach~\cite{Agui01,PRE-LV}.
Supposing that the collective behavior exhibited by biological systems can be parameterized by a Hamiltonian, $\mathcal{H}(x,\,k)$ with non-standard kinetic terms, quantum fluctuations could be identified and quantified through the WW QM~\cite{Novo2022,Novo2022A,Novo2024,Novo2025}.
Such an extended and equivalent description of the Schr\"odinger QM, encompasses the QM grounds expressed in terms of a {\em quasi}-probability distribution function spread over the phase-space~\cite{Wigner,Hillery,Ballentine,Case}.
Phase-space features of the Wigner flow for generic $1$-dim Hamiltonian constrained by $\partial^2 \mathcal{H} / \partial x \, \partial k=0$, where $x$ and $k$ are dimensionless position and momentum coordinates, then allows one to obtain Wigner functions and Wigner currents~\cite{Hillery,Zurek02,Steuernagel3} as a generic setup for the understanding of the quantization of competitive ecological systems~\cite{Novo2022,Novo2024}.

From a phenomenological perspective, prey-predator-like dynamics have already been studied in the context of competition-induced chaos. Results include, for instance, the understanding of symbiotic synchronization reproduced by {\em in vitro} molecular chains identified at biochemical and genomic systems~\cite{PP00,PP01,PP02,PP03,PP04}. Through bottom-up assembly of chemical components, Ref.~\cite{PP01} describes {\em in vitro} predator-prey-like oscillations where molecular behaviors emulating competition-induced chaos and symbiotic synchronization have been experimentally observed.
The potential connection between synthetic systems like these and prey-predator dynamics lies in the ability of such systems to exhibit self-organizing behavior, which can mimic ecological interactions. In biological ecosystems, predator-prey dynamics often lead to complex, cyclic behaviors driven by feedback loops between species. 
These synthetic systems are designed using a strategy that exploits the programmability of DNA interactions, precisely controlled by enzymatic catalysis. Once demonstrated experimentally, such systems can mirror the feedback-driven dynamics observed in predator-prey interactions, enhancing our understanding of molecular complexity. Moreover, they may enable the coordination of collective molecular behaviors in technological applications, similar to self-regulating ecological processes~\cite{PP01,PP03}.
On the theoretical side, unrestricted generalized frameworks for treating competitive ecological equilibrium of populations~\cite{RPSA-LV} still falls outside the context of variations of the Lotka-Volterra (LV) modified equations, for instance, as they are considered into the investigation of stochastic systems~\cite{RPSA-LV,Allen,Grasman}. 
These results suggest that ecosystem-like interaction complexities that affect the macroscopic features -- like stable coexistence, oscillations and chaos -- warrant further investigations from fundamental theoretical perspectives.

The WW QM framework indeed provides the tools for a self-contained statistical description of, at least, thermodynamic (TD) and Gaussian ensembles~\cite{Wigner,Hillery} in terms of Wigner flow and quantum information quantifiers obtained from corresponding Wigner currents, which account for the quantum fluctuations over some classical regime. For the above-mentioned class of constrained Hamiltonian systems, classicality and quantumness can be described through quantifiers of Liouvillianity and stationarity, in particular, in the specific context of prey-predator dynamics.

As an extension of the LV description~\cite{LV1,LV2,Novo2022,Novo2022B,Novo2022C}, which admits a dimensionless Hamiltonian formulation described by 
\begin{equation}\label{Ham}
\mathcal{H}(x,\,k)=a \,x + k + a\, e^{-x} + e^{-k},
\end{equation}
this work aims to perform a similar quantitative WW QM analysis for a Toda-like Hamiltonian expressed into the dimensionless form of~\cite{Novo2024}
\begin{equation}\label{Original}
\mathcal{H}_{\mbox{\tiny T}}(x,\,k)=\cosh(k) + a\,\cosh(x),
\end{equation}
in order to extend the investigation of the prey-predator competition dynamics in the framework of the WW QM~\cite{LV1,LV2,Novo2022,Novo2022B,Novo2024} and to quantify their quantum-related distortions.

The Hamiltonian in Eq.~(\ref{Original}), with $\cosh(x)$ replaced by generic values of a position-dependent potential, $\mathcal{V}(x)$, has appeared in the eigenvalue problem that leads to a difference equation similar to the Baxter equation of the Toda lattice~\cite{Este,90}.
In this context, the quantum Toda lattice has been solved through the so-called Baxter correspondence~\cite{89}, providing solution perspectives to the Baxter equation~\cite{90}.
For the general form $\mathcal{H}_{\mbox{\tiny T}}(x,\,k)=\cosh(k) + \mathcal{V}(x)$, it has also been related to the quantization of the Seiberg-Witten curve in $N=2$ Yang-Mills theory with gauge group SU(N)~\cite{95}, which again reproduces the spectral decomposition of the periodic Toda lattice~\cite{46,80}.
In more complex scenarios, exact quantization conditions for the resulting spectral problem depicted by mirror curves of toric Calabi-Yau manifolds were also conjectured~\cite{21,49}.
The encompassing discussion of these models, where a framework to obtain exact quantization conditions and the actual spectral decomposition for the Hamiltonian (\ref{Original}) has been identified, was investigated in Ref.\cite{Este} in the context of a solvable deformation of QM.
Despite its demonstrated connection with ordinary QM, Hamiltonians like (\ref{Original}) suggest that an analytic determination of the scaling regime is still under investigation.
Therefore, besides the classical appeal related to the prey-predator dynamics, the gap in understanding the QM of (\ref{Original}) can be partially addressed by the framework considered here.

Considering that the WW formalism~\cite{Wigner,Ballentine,Case} -- once extended to a specific class of non-standard Hamiltonians described by $\mathcal{H}(x,\,k)=\mathcal{K}(k) + \mathcal{V}(x)$, with arbitrary functions $\mathcal{K}(k)$ and $\mathcal{V}(x)$~\cite{Novo2022A,Novo2022B,Novo2024,Novo2025} -- allows for a systematic understanding of the role of quantum effects on the classical background, our work suggests that the Toda-like Hamiltonian (\ref{Original}) can support a description (cf. Refs.~\cite{Novo2022,NossoPaper,NossoPaper19,Meu2018}) of species' phase-space correlations and information flow aspects, as well as their quantum-like deviations from classical patterns.

Assuming that classical and quantum evolutions coexist at different scales, macroscopic (classical) and microscopic (quantum) patterns for such Hamiltonians are both affected by the influence of the magnitude of the Planck constant, $\hbar$.
Mapping the effective canonical non-commutative residue of $x$ and $k$ operations, $[x,\,k] \neq 0$, into the prey-predator system implies that quantum-like effects over the averaged values of $x$ and $k$ can be depicted in a macroscopic frame\footnote{This can be identified for dynamical systems where either physical or phenomenological constraints produce non-commutativity between canonical variables as an emergent property. For instance, for the Aubry(--Andr\'e)--Harper dynamics~\cite{Novo2022A,Novo2025}, one has $[x,\,k] = i\,2\pi\beta$, where the so-called Peierls phase parameter, $\beta$, plays the role of an effective Planck constant.}.

As performed for LV Hamiltonian systems, the quantization paradigm involving the Toda-like Hamiltonian from Eq.~(\ref{Original}) may serve to explain (classical) collective behaviors~\cite{Bio17}, correlated with temporal and spatial (state) descriptions of species distributions, which may admit some {\em quantum-analog} non-commutative property of $x$ and $k$, i.e., $[x,\,k] \neq 0$, reflected in the observable outcomes.

The outline of our manuscript is as follows.
Sec.~II addresses the classical features driven by the Toda-like Hamiltonian from Eq.~(\ref{Original}), the perturbative evaluation of TD ensembles and the exact evaluation of Gaussian ones.
The results include a precise evaluation of stationarity conditions in the context of classical-to-quantum correspondence, as well as the connection with statistical mechanics.
Specifically, for Gaussian ensembles, the overall quantum distortion pattern is computed non-perturbatively in terms of a convergent infinite series expansion involving quadratic powers of the Planck constant, $\hbar$.
In Sec.~IV, the semiclassical dynamical patterns for Gaussian ensembles and the corresponding quantum-distorted prey-predator phase-space trajectories are obtained.
Our conclusions are drawn in Sec.~V to connect with some imprints from quantumness and stability of prey-predator dynamics previously investigated.

\section{Classical features}

To understand the naive correspondence between Toda-like Hamiltonians in the form of Eq.~\eqref{Original} and the phenomenology of prey-predator dynamics at a classical level, one can depart from the well-known 2-dim LV system of equations put into the form of
\begin{eqnarray}\label{Ham2bbbA}
d{y}/d\tau &=& y\,z - y,\\
\label{Ham2bbbB}
d{z}/d\tau &=& a\,(z - y\,z),
\end{eqnarray}
where $y(\tau)$ and $z(\tau)$ correspond to the predator and prey populations, respectively normalized by their corresponding time-averaged mean populations which set equilibrium points at $y=z=1$.

The solutions for Eqs.~\eqref{Ham2bbbA}-\eqref{Ham2bbbB} correspond to $y-z$ phase-space implicit level curves, the solution curves $\Gamma_{{H}=\epsilon}$, captured by
\begin{eqnarray}\label{Ham2ccc}
{H}(y,\, z)=a\,y + z - \ln(y^a\,z),
\end{eqnarray}
with an arbitrary anisotropy parameter, $a$, and $\epsilon$ assuming constant values constrained by the condition $\epsilon > a+1$.

Semi-analytical solutions for Eqs.~\eqref{Ham2bbbA}-\eqref{Ham2bbbB} are only admitted for the isotropic coordinate version of the system (i.e. with $a=1$), with the phase-space trajectories identified by the parametric curve,
\begin{eqnarray}\label{Ham2B}
 y(\mathcal{T})&=&\frac{\mathcal{T} \pm \sqrt{\mathcal{T}^2-4 e^{\mathcal{T} - \epsilon}}}{2},\\
 z(\mathcal{T})&=&\frac{\mathcal{T} \mp \sqrt{\mathcal{T}^2-4 e^{\mathcal{T} - \epsilon}}}{2},
\end{eqnarray}
with the dynamical constraint,
\begin{eqnarray}\label{Ham2C}
\dot{\mathcal{T}}^2 - {\mathcal{T}^2-4 e^{\mathcal{T} - \epsilon}}=0,
\end{eqnarray}
where ``{\em dots}'' correspond to $\tau$ derivatives, $d/d\tau$, and $\epsilon$ is identified as a (dimensionless energy) constant of motion.

\begin{figure}[h]
\includegraphics[scale=0.8]{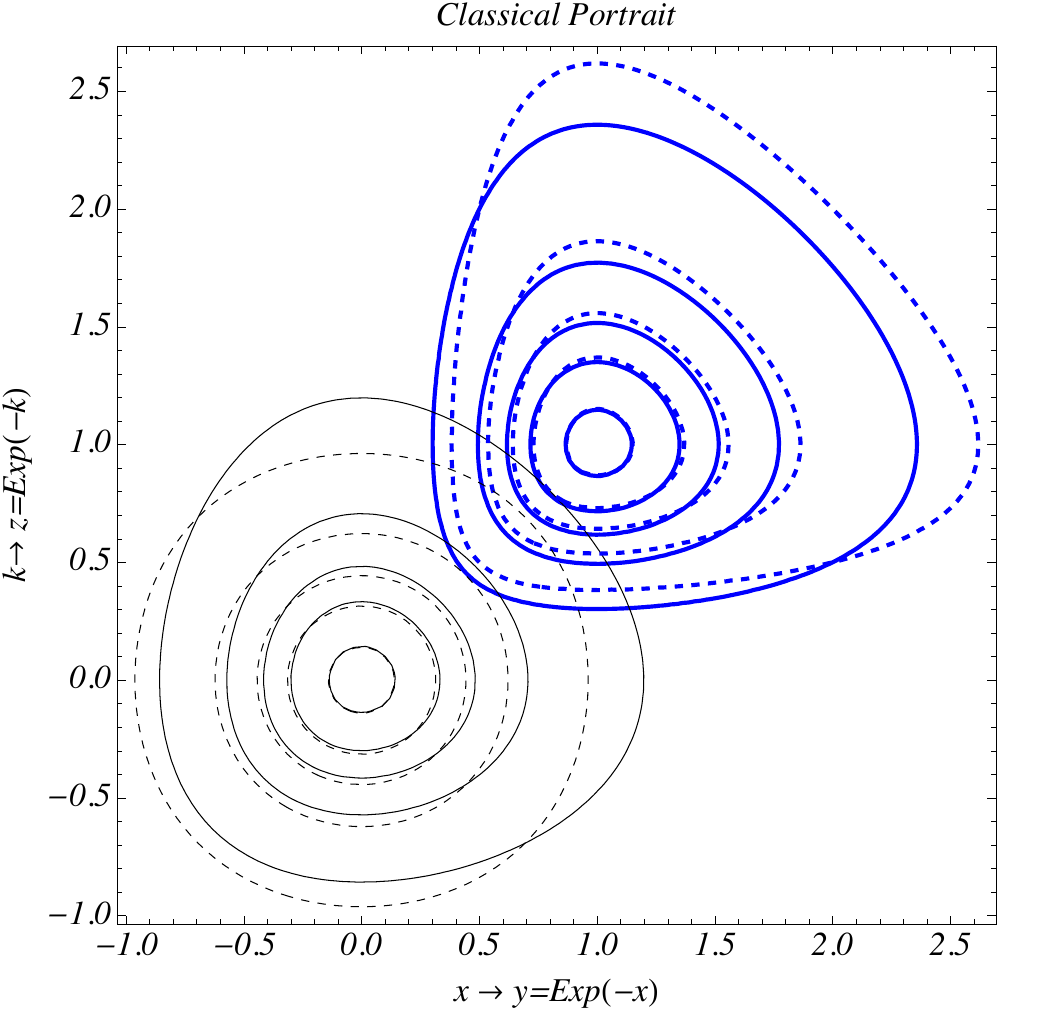}
\renewcommand{\baselinestretch}{.65}
\caption{\footnotesize
(Color online) Classical portrait of LV (solid lines) and Toda-like (dashed lines) associated Hamiltonians. The phase-space $x-k$ trajectories (black thin lines) are for $\epsilon= 2.5,\, 2.2,\, 2.1,\, 2.05$ and $2.01$, and the corresponding species distributions (red thick lines), $z$ and $y$, are identified by $y=e^{-x}$ and $z=e^{-k}$. The values of $\epsilon$ are chosen to denote the consistence with Eq.~\eqref{HamBm}.}
\label{LVLVLV}
\end{figure}

The singular mathematical manipulability properties of LV equations~\cite{RPSA-LV,Allen,Grasman} can be encompassed by a dimensionless Hamiltonian formulation given in terms of Eq.~\eqref{Ham},
which yields classical equations of motion written in terms of the Poisson brackets (PB) as
\begin{eqnarray}\label{Ham2s}
d{x}/d\tau &=& \{x,\mathcal{H}\}_{PB}=1-e^{-k},\\
\label{Ham2t}d{k}/d\tau &=& \{k,\mathcal{H}\}_{PB}=a\,e^{-x} - a,
\end{eqnarray}
which, by the way, describe $x-k$ oscillations as a natural mode of population coexistence in ecological chains, with the number of species, $y$ and $z$, identified by $y=e^{-x}$ and $z=e^{-k}$, as depicted in the phase-space maps (solid lines) from Fig.~\ref{LVLVLV}.

Turning back to the Toda-like Hamiltonian, Eq.~\eqref{Original} exhibits a similar structure, which is reflected by classical equations of motion in the form of
\begin{eqnarray}\label{Ham2Bs}
d{x}/d\tau &=& \{x,\mathcal{H}_{\mbox{\tiny T}}\}_{PB}=\sinh(k),\\
\label{Ham2Bt}d{k}/d\tau &=& \{k,\mathcal{H}_{\mbox{\tiny T}}\}_{PB}=-a\,\sinh(x),
\end{eqnarray}
which maps $x-k$ oscillations from Eqs.~\eqref{Ham2s}-\eqref{Ham2t} up to quadratic order of $x$ and $k$ expansions. In fact, one can notice that 
\begin{equation}\label{HamBm}
\mathcal{H}(x,\,k)=(1+a) + \frac{1}{2}\left(x^2 + k^2)\right) + \mathcal{O}(x^3,\,k^3)=\mathcal{H}_{\mbox{\tiny T}}(x,\,k).
\end{equation}
The prey-predator-like map, $y=e^{-x}$ and $z=e^{-k}$, is thus straightforwardly identified by setting the constraint 
\begin{eqnarray}\label{Ham2cccB}
{H}_{\mbox{\tiny T}}(y,\, z)=\frac{1}{2}\left(a\,y + \frac{a}{y} + z + \frac{1}{z}\right)=\epsilon,
\end{eqnarray}
with the corresponding 2-dim system obtained from Eqs.~\eqref{Ham2Bs}-\eqref{Ham2Bt} written as
\begin{eqnarray}\label{Ham2bbb22w}
d{y}/d\tau &=& \frac{1}{2}\,\left(y\,z-\frac{y}{z} \right),\\
\label{Ham2bbb22t}d{z}/d\tau &=& \frac{a}{2}\,\left(\frac{z}{y}-y\,z\right),
\end{eqnarray}
from which, for $\Gamma_{{H}_{\mbox{\tiny T}}=\epsilon}$ phase-space closed paths (as depicted in the phase-space maps (dashed lines) from Fig.~\ref{LVLVLV}), one has
\begin{equation}
\int_0^{T}d\tau\, y^{-1} (d{y}/d\tau)=\int_0^{T}d\tau\, z^{-1} (d{z}/d\tau)=0,
\end{equation}
with a periodic ciclic dynamics identified by the period of time, $T$, computed from a direct $\tau$ integration of Eq.~\eqref{Ham2cccB},
\begin{eqnarray}
\frac{1}{\epsilon}\int_0^{T}d\tau\, (ay+z)=\frac{1}{\epsilon}\int_0^{T}\,d\tau\, (ay^{-1}+z^{-1})=T,
\end{eqnarray}
with auxiliary coordinate integrals evaluated as
\begin{eqnarray}\label{Ham2bbb33s}
\int_0^{T}d\tau\, y &=& \int_0^{T}d\tau\, y^{-1},\\
\label{Ham2bbb33t}\int_0^{T}d\tau\, z &=& \int_0^{T}d\tau\, z^{-1}.
\end{eqnarray}

\begin{figure}[h]
\includegraphics[scale=0.8]{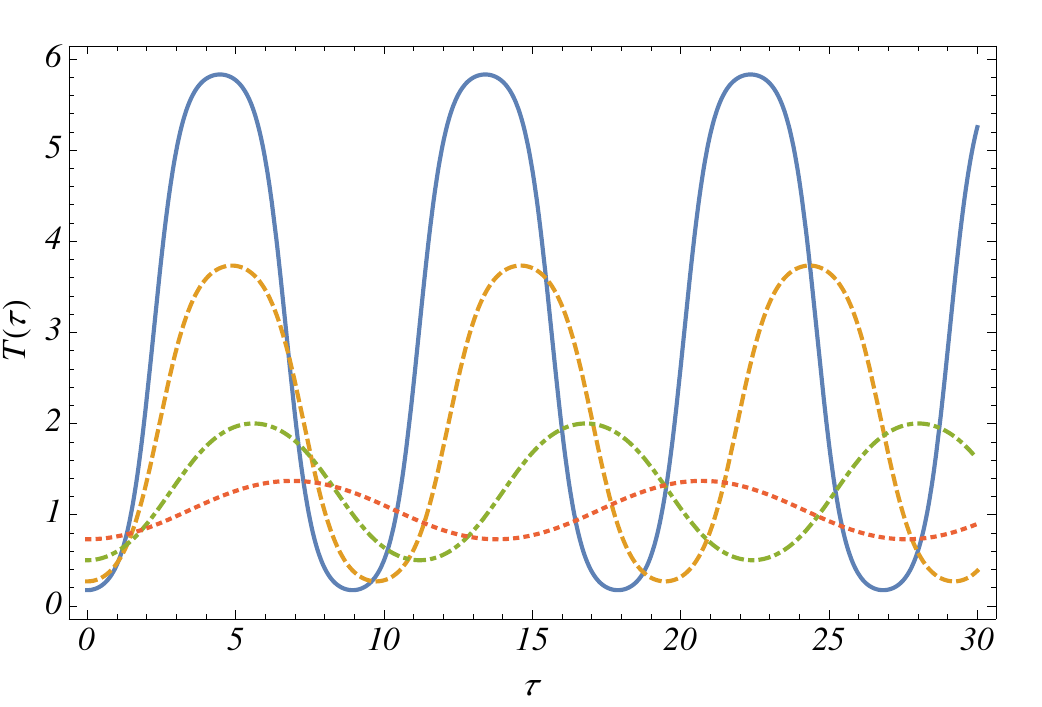}
\renewcommand{\baselinestretch}{.65}
\caption{\footnotesize
(Color online) Time dependence of the parameter $\mathcal{T}(\tau)$ for the classical Toda-like dynamical system. Results are are for $\epsilon=6,\, 4,\, 2.5$ and $2.1$.}
\label{LVTodaMod}
\end{figure}

From Fig.~\ref{LVLVLV}, one can notice that, for smaller oscillation amplitudes, an approximated harmonic oscillation pattern (acording to Eq.~\eqref{HamBm}) is identified in both descriptions. For the identical values of the energy parameter, $\epsilon$, the Toda-like formulation leads to increasing oscillation amplitudes with respect to the LV original formulation. This suggests a suitable modification in the prey-predator dynamics.
From a straight mathematical perspective, besides the parity symmetry now exhibited by the Toda-like Hamiltonian map, analytical solutions can be obtained for the system $y-z$ isotropic configuration of Eqs.~\eqref{Ham2bbb33s}-\eqref{Ham2bbb33t}, as they lead to
\begin{eqnarray}\label{sols}
y(\tau) &=&\mathcal{T} \pm \sqrt{\mathcal{T}^2-\frac{\mathcal{T}}{\epsilon-\mathcal{T}}},\nonumber\\
z(\tau) &=&\mathcal{T} \mp \sqrt{\mathcal{T}^2-\frac{\mathcal{T}}{\epsilon-\mathcal{T}}}.
\end{eqnarray}
The dynamical constraint identified by
\begin{eqnarray}\label{Ham2CBB}
\dot{\mathcal{T}}^2 - \mathcal{T}^2(\mathcal{T} - \epsilon)^2-\mathcal{T}(\mathcal{T} - \epsilon)=0,
\end{eqnarray}
now results into
\begin{eqnarray}\label{Ham2CBBB}
\mathcal{T}(\tau) &=& \frac{2}{\sqrt{\epsilon^2-4} \left(1-2 \,\textit{sn}\left(\frac{ \sqrt{\left(\epsilon+\sqrt{\epsilon^2-4}\right)-2}}{2 \sqrt{2}}\tau\bigg{\vert}\frac{2 \epsilon \sqrt{\epsilon^2-4}}{\epsilon \left(\epsilon+\sqrt{\epsilon^2-4}\right)-2}\right)^2\right)+\epsilon},
\end{eqnarray}
where $\textit{sn}(u|\kappa)$ are the Jacobian elliptic function with the elliptic modulus,
$$\kappa \equiv \kappa(\epsilon)=\frac{2 \epsilon \sqrt{\epsilon^2-4}}{\epsilon \left(\epsilon+\sqrt{\epsilon^2-4}\right)-2},$$
for which the $\mathcal{T}(\tau)$ oscillatory behavior is be depicted in Fig.~\ref{LVTodaMod}.

In particular, one notices that the $\mathcal{T}(\tau)$ oscillation is bounded by 
$$\mathcal{T}_{\pm}=\frac{1}{2}(\epsilon\pm\sqrt{\epsilon^2-4})$$
upper ($+$) and lower ($-$) limits, which coincide with prey and predator amplitude limits, $y_{\pm}=\mathcal{T}_{\pm}=z_{\pm}$, and oscillates with a period, $T(\epsilon)$, given by
\begin{equation}\label{period}
T(\epsilon)=\frac{8\sqrt{2}K(\kappa(\epsilon))}{ \sqrt{\left(\epsilon+\sqrt{\epsilon^2-4}\right)-2}},\end{equation}
that, on its turn, follows from the period of the Jacobian elliptic function, $\textit{sn}(...)$, straightforwardly obtained from complete elliptic integral of the first kind,
\begin{eqnarray}\label{period}
K(\kappa(\epsilon)) &=& 4 \int_{0}^{\pi/2} d\theta \left[1-\kappa(\epsilon)\,\sin(\theta)\right]^{-1/2},
\end{eqnarray}
modulated by the multiplying factor $\frac{ \sqrt{\left(\epsilon+\sqrt{\epsilon^2-4}\right)-2}}{2 \sqrt{2}}$ in the argument of 
$\textit{sn}(...)$.

Despite the explicit diffeomorphism between the symplectic manifolds (tori) of the LV system and the one-particle Toda lattice -- which can indeed be identified for any two Hamiltonians written in the form $\mathcal{H}(x,\,k)=\mathcal{K}(k)+\mathcal{V}(x)$, provided that $\mathcal{K}(k)$ and $\mathcal{V}(x)$ are invertible functions\footnote{As obtained from Eqs.~\eqref{Ham2ccc} and \eqref{Ham2cccB}, with the replacements of $y=e^{-x}$ and $z=e^{-k}$.} -- the above results point to inequivalent outcomes, at least from the perspective of obtaining analytical solutions for classical and quantum observables.

For LV systems, the period $T$ cannot be expressed explicitly in terms of $\varepsilon$, since Eq.~\eqref{Ham2C} is not analytically integrable. Nevertheless, a Bohr-Sommerfeld quantization rule can be parameterized~\cite{Novo2024}.
By contrast, for the Toda-like systems discussed above, even though the period $T$ can be computed in terms of $\varepsilon$, an equivalent Bohr-Sommerfeld quantization expressed in terms of the canonical variables $x$ and $k$ cannot be evaluated analytically.
However, in both cases, to investigate the corresponding quantum features, one can follow the WW framework applied to non-standard Hamiltonian systems of the form $\mathcal{H}(x,\,k)=\mathcal{K}(k)+\mathcal{V}(x)$.

\section{WW quantum features}

The Wigner function, $W(q,\, p)$, written in terms of physical coordinates of position and momentum, $q$ and $p$, is the functional statistical tool which drives the WW formulation of QM and encompasses
its paradigms in terms of a {\em quasi}-probability distribution function of $q$ and $p$~\cite{Wigner,Hillery,Ballentine,Case}.
Such a phase-space extended and equivalent description of the Schr\"odinger QM (see the Appendix) includes the interpretation of phase-space features in therms of fluid-analogs related to the Wigner flow properties described by Wigner currents.

In such a context, the dynamical evolution of the Wigner function, $W(q,\,p) \to W(q,\,p;\,t)$, arises from the continuity equation~\cite{Case,Ballentine,Steuernagel3,NossoPaper,NossoPaper19,Meu2018},
\begin{equation}
{\partial_t W} + {\partial_q J_q}+{\partial_p J_p} =0,\qquad \mbox{with}\quad \partial_a \equiv \partial/\partial a,
\label{z51}
\end{equation}
with the flow field described in terms of a vector flux, $\mathbf{J}=J_q\,\hat{q} + J_p\,\hat{p}$, whose the corresponding phase-space components are written as~\cite{Steuernagel3,NossoPaper,NossoPaper19,Meu2018,Novo2022,Novo2022B}
\begin{equation}
J_q(q,\,p;\,t)=+\sum_{\eta=0}^{\infty} \left(\frac{i\,\hbar}{2}\right)^{2\eta}\frac{1}{(2\eta+1)!} \, \left[\partial_p^{2\eta+1} K(p)\right]\,\partial_q^{2\eta}W(q,\,p;\,t),
\label{z500BB}
\end{equation}
and
\begin{equation}
J_p(q,\,p;\,t)=-\sum_{\eta=0}^{\infty} \left(\frac{i\,\hbar}{2}\right)^{2\eta}\frac{1}{(2\eta+1)!} \, \left[\partial_q^{2\eta+1} V(q)\right]\,\partial_p^{2\eta}W(q,\,p;\,t),\label{z500CC}
\end{equation}
where $K(p)$ and $V(q)$ are, respectively, generic $p$ and $q$ dependent contributions to the Weyl transformed Hamiltonian, ${H}^W(q,\,p)=K(p) + V(q)$, constrained by the condition $\partial^2 {H}^W / \partial q \, \partial p=0$. 

The quantum contributions which affect the evolution of $W(q,\,p;\,t)$ driven by ${H}^W(q,\,p)$ can be constructed through the series expansion contributions from $\eta \geq 1$. Once they are suppressed, the classical Hamiltonian equations are recovered, 
\begin{equation}
J^{\mathcal{C}}_q(q,\,p;\,t)= +({\partial_p H})\,W(q,\,p;\,t), \label{z500BB2}
\end{equation}
and
\begin{equation}
J^{\mathcal{C}}_p(q,\,p;\,t)=-({\partial_q H})\,W(q,\,p;\,t),
\label{z500CC2}
\end{equation}
with the upper index ``$^W$'' for $H$ being suppressed from now on.

Quantum distortions over the classical Hamiltonian regime can be more conveniently manipulated with related variables recasted in terms of dimensionless variables, $x=\left(m\,\omega\,\hbar^{-1}\right)^{1/2} q$ and $k=\left(m\,\omega\,\hbar\right)^{-1/2}p$, where $m$ is a mass scale parameter and $\omega$ is an arbitrary angular frequency.
In this case, a dimensionless Hamiltonian is written as $\mathcal{H}(x,\,k)=\mathcal{K}(k) + \mathcal{V}(x)$, in correspondence with $\mathcal{H}=(\hbar \omega)^{-1} H$, with $\mathcal{V}(x)=(\hbar \omega)^{-1} V\left(\left(m\,\omega\,\hbar^{-1}\right)^{-1/2}x\right)$ and $\mathcal{K}(k)=(\hbar \omega)^{-1} K\left(\left(m\,\omega\,\hbar\right)^{1/2}k\right)$. Similarly, the Wigner function can be recast in the form of $\mathcal{W}(x, \, k;\,\tau) \equiv \hbar\, W(q,\,p;\,t)$, 
\begin{eqnarray}\label{xDimW}
\mathcal{W}(x, \, k;\,\tau) &=& \pi^{-1} \int^{+\infty}_{-\infty} \hspace{-.35cm}du\,\exp{\left(2\, i \, k \,u\right)}\,\phi(x - u;\,\tau)\,\phi^{\ast}(x + u;\,\tau),
\end{eqnarray}
with $\hbar$ absorbed by the phase-space volume $dp\,dq\to \hbar\, dx\,dk$.
At Eq.~\eqref{xDimW}, one identifies $u=\left(m\,\omega\,\hbar^{-1}\right)^{1/2} s$, and $\tau=\omega t$, such that the dimensionless Wigner currents turn into the form of~\cite{Novo2022,Novo2022B}
\begin{eqnarray}
\label{imWA}\mathcal{J}_x(x, \, k;\,\tau) &=& +\sum_{\eta=0}^{\infty} \left(\frac{i}{2}\right)^{2\eta}\frac{1}{(2\eta+1)!} \, \left[\partial_k^{2\eta+1}\mathcal{K}(k)\right]\,\partial_x^{2\eta}\mathcal{W}(x, \, k;\,\tau),\\
\label{imWB}\mathcal{J}_k(x, \, k;\,\tau) &=& -\sum_{\eta=0}^{\infty} \left(\frac{i}{2}\right)^{2\eta}\frac{1}{(2\eta+1)!} \, \left[\partial_x^{2\eta+1}\mathcal{V}(x)\right]\,\partial_k^{2\eta}\mathcal{W}(x, \, k;\,\tau),
\end{eqnarray}
which render the continuity equation in a dimensionless form,
\begin{equation}\label{z51dim}
{\partial_{\tau} \mathcal{W}} + {\partial_x \mathcal{J}_x}+{\partial_k \mathcal{J}_k}={\partial_{\tau} \mathcal{W}} + \mbox{\boldmath $\nabla$}_{\xi}\cdot\mbox{\boldmath $\mathcal{J}$} =0.
\end{equation}

From Eqs.~\eqref{imWA} and \eqref{imWB}, one notices that quantum corrections are highly affected by entangled non-linear contributions involved in the series expansion.
Therefore, in order to distinguish quantum from classical regimes, as well as to identify their corresponding stability conditions, the still entangled stationarity quantifier
\begin{equation} \label{helps}
\partial_{\tau} \mathcal{W}= -\mbox{\boldmath $\nabla$}_{\xi}\cdot\mbox{\boldmath $\mathcal{J}$} =\sum_{\eta=0}^{\infty}\frac{(-1)^{\eta}}{2^{2\eta}(2\eta+1)!} \, \left\{
\left[\partial_x^{2\eta+1}\mathcal{V}(x)\right]\,\partial_k^{2\eta+1}\mathcal{W}
-
\left[\partial_k^{2\eta+1}\mathcal{K}(k)\right]\,\partial_x^{2\eta+1}\mathcal{W}
\right\},\end{equation}
requires complementary quantitative information tools~\cite{Novo2022,Novo2022B}.
In particular, local quantum distortions can be quantified by the differential operator defined by~\cite{Novo2022,Novo2022B}
\begin{equation}
\mbox{\boldmath $\nabla$}_{\xi} \cdot \mathbf{w}=\sum_{\eta=0}^{\infty}\frac{(-1)^{\eta}}{2^{2\eta}(2\eta+1)!}
\left\{
\left[\partial_k^{2\eta+1}\mathcal{K}(k)\right]\,
\partial_x\left[\frac{1}{\mathcal{W}}\partial_x^{2\eta}\mathcal{W}\right]
-
\left[\partial_x^{2\eta+1}\mathcal{V}(x)\right]\,
\partial_k\left[\frac{1}{\mathcal{W}}\partial_k^{2\eta}\mathcal{W}\right]
\right\},~~~ \label{zeqnz59ss}
\end{equation}
with $\mathbf{w}= \mbox{\boldmath $\mathcal{J}$}/\mathcal{W}$, where deviations from classical patterns can be identified by a kind of non-Liovillian condition given by $\mbox{\boldmath $\nabla$}_{\xi} \cdot \mathbf{w}\neq 0$.
The meaning of Liouvillianity in such a context has already been broadly discussed in the literature~\cite{Case,Ballentine,Steuernagel3,NossoPaper,NossoPaper19,Meu2018,Novo2022,Novo2022B}\footnote{The Liouville equation can be recovered from Eq.~\eqref{z51dim} by replacing the corresponding Wigner currents, $\mathcal{J}_x$ and $\mathcal{J}_k$, by the corresponding dimensionless forms of those from Eqs.~\eqref{z500BB2} and \eqref{z500CC2}, respectively.}.

In this case, the above-introduced quantity, $\mathbf{w}$, is replaced by the phase-space velocity along the classical trajectory, $\mathcal{C}$, which is identified by $\mathbf{v}_{\xi(\mathcal{C})}=\dot{\mbox{\boldmath $\xi$}}=(\dot{x},\,\dot{k})\equiv ({\partial_k \mathcal{H}},\,-{\partial_x \mathcal{H}})$, and exhibits the divergenceless behavior, $\mbox{\boldmath $\nabla$}_{\xi}\cdot \mathbf{v}_{\xi(\mathcal{C})}= \partial_x \dot{x} + \partial_k\dot{k}=0$, as the imprint of the Liouvillian regime.
The quantum current associated velocity implicitly given in terms of $\mbox{\boldmath $\mathcal{J}$}=\mathbf{w}\,\mathcal{W}$, leads to the interpretation of $\mathbf{w}$ as the {\em quantum analog} of $\mathbf{v}_{\xi(\mathcal{C})}$, i.e. $\mathbf{v}_{\xi(\mathcal{C})}$ is the semi-classical limit ($\hbar \to 0$) of $\mathbf{w}$.
In this case, $\mbox{\boldmath $\nabla$}_{\xi}\cdot\mbox{\boldmath $\mathcal{J}$}=\mathcal{W}\,\mbox{\boldmath $\nabla$}_{\xi}\cdot\mathbf{w}+ \mathbf{w}\cdot \mbox{\boldmath $\nabla$}_{\xi}\mathcal{W}$ is more elegantly written as~\cite{Steuernagel3}
\begin{equation}
\mbox{\boldmath $\nabla$}_{\xi} \cdot \mathbf{w}=\frac{\mathcal{W}\, \mbox{\boldmath $\nabla$}_{\xi}\cdot \mbox{\boldmath $\mathcal{J}$} - \mbox{\boldmath $\mathcal{J}$}\cdot\mbox{\boldmath $\nabla$}_{\xi}\mathcal{W}}{\mathcal{W}^2},
\label{zeqnz59}
\end{equation}
and the quantifiers for stationary and Liouvillian regimes, $\mbox{\boldmath $\nabla$}_{\xi}\cdot\mbox{\boldmath $\mathcal{J}$}$ and $\mbox{\boldmath $\nabla$}_{\xi} \cdot \mathbf{w}$, provide the expected information content.

Additional effective tools for quantifying the flux of information through the phase-space parametric classical surface, $\mathcal{C}$, have also been considered to describe quantum systems with clearly identified spectral patterns~\cite{NossoPaper,NossoPaper19}. Since this is not the case of the solutions driven by the Hamiltonian Eq.~\eqref{Original}, one should then consider that stationarity and stability properties can be perturbatively and non-perturbatively computed for statistical quantum ensembles of associated with TD and Gaussian natures, respectively~\cite{Novo2022}.

As depicted in the following, once such a statistical description of a Toda-like Hamiltonian is provided, the quantum fluctuations over the classical regime can be quantified in terms of the Wigner flow features. In particular, the concepts of stationarity and quantumness, quantified in terms of Eqs.~\eqref{helps} and \eqref{zeqnz59ss}, respectively, shall be discussed in the context of the dynamics of prey-predator systems now parameterized by the Toda-like Hamiltonian such that some phenomenological issues related to non-extinction/discreteness and stability properties~\cite{Novo2022,Novo2022B,Novo2022C} can then be addressed through the comparison between classical and quantum patterns.

\subsection{Perturbative evaluation of TD ensembles}

The classical-to-quantum correspondence involving canonical ensembles is related to the phase-space probability and information flow interpretations devised by Wigner~\cite{Wigner, Hillery}, which stablish the correspondence between classical mechanics and classical TD on the same level of that between QM and quantum TD.

The statistical properties of classical ensembles are therefore described by the Maxwell-Boltzmann distribution, which can be cast in the form of
\begin{equation}\label{TDclass}
\mathcal{W}_0(x,\,k;\,\beta)=\mathcal{Z}^{-1}_0(\beta)\,\exp[-\beta\,\mathcal{H}(x,\,k)],
\end{equation}
with the partition function, $\mathcal{Z}_0(\beta)$, given by
\begin{equation}
\mathcal{Z}_0(\beta)=
\int^{+\infty}_{-\infty} \hspace{-.35cm}dx\,\int^{+\infty}_{-\infty} \hspace{-.35cm}dk\,\exp[-\beta\,\mathcal{H}(x,\,k)],
\end{equation}
where the dimensionless parameter, $\beta$, sets the dependence on the temperature, $\mathcal{T}$, by $\beta=\hbar \omega/ k_{B} \mathcal{T}$, where $k_{B}$ is the Boltzmann's constant.
The quantized form of $\mathcal{W}_0(x,\,k;\,\beta)$ is given in terms of the sum $\sum_{\ell}\exp(- E_{\ell}/k_B\mathcal{T}) \mathcal{W}_{\ell}(x,\, k)$, where the {\em eigen}energies, $E_{\ell}$, are discretized by the quantum number $\ell$, in correspondence with WW transformed stationary {\em eigen}functions, $\mathcal{W}_{\ell}(x,\, k)$.
However, for most cases, a spectral decomposition of the quantum system in terms of $\{E_{\ell}\}$ {\em eigen}values is not analytically admitted. Quantum corrections to the classical TD regime can then be obtained through a perturbative procedure.
From stationary solutions for the Wigner probability flux cf. Eq.~\eqref{z51dim}, $\mathcal{W}_{ST}(x,\, k;\,\beta)$, and searching for an approximation to the form of $\sum_{\ell}\exp(- E_{\ell}/k_B\mathcal{T}) \mathcal{W}_{\ell}(x,\, k)$, up to order $\mathcal{O}(\hbar^{2N})$, one can identify the perturbative quantum modifications from the series expansion given by~\cite{Wigner,Coffey07,Novo2022}
\begin{equation} \label{serrs}
W^{(2N)}_{ST}(x,\, k;\,\beta)=\sum_{\eta=0}^N\,\hbar^{2\eta}\,\mathcal{W}_{2\eta}(x,\, k;\,\beta),
\end{equation}
which corresponds to the closed form of the equilibrium equation introduced by Wigner~\cite{Wigner}. 
Such a Wigner prescription can be interpreted as an adiabatically driven mechanism through which TD equilibrium conditions are achieved as a response of the quantum system to the quantum perturbations in order to result into the Wigner stationary distribution in the form of the limit $\mathcal{W}_{ST}(x,\, k;\,\beta)=\lim_{N\to \infty}W^{(2N)}_{ST}(x,\, k;\,\beta)$.

Through an approach where contributions from $\mathcal{O}(\hbar^{4})$ or higher can be suppressed, the Wigner strategy has been extended~\cite{Novo2022,Novo2022B} to Hamiltonians in the arbitrary form of $\mathcal{H}(x,\,k)=\mathcal{K}(k) + \mathcal{V}(x)$.

With $\mathcal{W}_{2}(x,\, k;\,\beta)$ identified by $\varepsilon_{(x,\, k;\,\beta)}\, \mathcal{W}_{0}(x,\, k;\,\beta)$, the procedure from Refs.~\cite{Novo2022,Novo2022B} leads to
\begin{eqnarray}\label{oideem}
\mathcal{W}^{(2)}_{ST}(x,\,k;\,\beta) &=& \frac{\mathcal{Z}_0(\beta)}{ \mathcal{Z}_{ST}(\beta)}\,\mathcal{W}_{0}(x,\,k;\,\beta)
\left[
1 +\varepsilon_{(x,\,k;\,\beta)}\right],
\end{eqnarray}
with
\begin{eqnarray}
\varepsilon_{(x,\,k;\,\beta)} &=&
-\frac{\beta^2}{8}
\partial_x^{2}\mathcal{V}(x)\,\partial^2_k \mathcal{K}(k)
+\frac{\beta^3}{24}
\left[
\partial_x^{2}\mathcal{V}(x)\,\left(\partial_k \mathcal{K}(k)\right)^2
+
\partial_k^{2}\mathcal{K}(k)\,\left(\partial_x \mathcal{V}(x)\right)^2
\right],
\end{eqnarray}
where the normalization pre-factor, ${\mathcal{Z}_0(\beta)}/{ \mathcal{Z}_{ST}(\beta)}$, has been introduced to normalize the associated probabilities, and the $\mathcal{O}(\hbar^{2})$ perturbative contributions were converted into dimensionally equivalent $\mathcal{O}\left(\beta^3\right)$ corrections, given that, in both cases, $\mathcal{O}(\hbar^{4})$ and $\mathcal{O}(\beta^{4})$ contributions are suppressed.

As from Refs.~\cite{Novo2022,Novo2022B}, the Wigner current components are straightforwardly written as
\begin{equation}
\mathcal{J}^{(2)}_x(x,\, k;\,\tau)=+ \left\{ \partial_k \mathcal{K}(k) \left(1 + \varepsilon_{(x,\, k;\,\beta)}\right) 
- \frac{1}{24}\,\partial_k^3 \mathcal{K}(k)\,
\left[\beta^2 \left(\partial_x \mathcal{V}(x)\right)^2 -\beta \partial^2_x \mathcal{V}(x)\right]\right\}\mathcal{W}_0,
\label{z500BB}
\end{equation}
and
\begin{equation}
\mathcal{J}^{(2)}_k(x,\, k;\,\tau)=- \left\{ \partial_x \mathcal{V}(x) \left(1 + \varepsilon_{(x,\, k;\,\beta)}\right) 
- \frac{1}{24}\,\partial_x^3 \mathcal{V}(x)\,
\left[\beta^2 \left(\partial_k \mathcal{K}(k)\right)^2 -\beta \partial^2_k \mathcal{K}(k)\right]\right\}\mathcal{W}_0,
\label{z500BB}
\end{equation}
and, up to $\mathcal{O}(\hbar^2)$, the modifications on the Liouvillian pattern are identified by the non-vanishing value of 
\begin{equation}
\mbox{\boldmath $\nabla$}_{\xi} \cdot \mathbf{w}=\frac{\beta^2}{12}
\left[
\partial_k^3 \mathcal{K}(k)\partial^2_x \mathcal{V}(x)\partial_x \mathcal{V}(x)
-
\partial_x^3 \mathcal{V}(x)\partial^2_k \mathcal{K}(k)\partial_k \mathcal{K}(k)\right],
\end{equation}
which has also been written into a dimensionless form.

Given the above considerations, for $\mathcal{K}(k)=\cosh(k)$ and $\mathcal{V}(x)= a \, \cosh(x)$, as identified from the Hamiltonian Eq.~\eqref{Ham}, one thus write the Toda-like associated classical distribution as
\begin{equation}
\mathcal{W}_0(x,\,k;\,\, \beta)=\mathcal{Z}^{-1}_0(\beta)\,\exp\left\{-\, \beta \left[a\,\cosh(x) + \cosh(k)\right]\right\},
\end{equation}
with the phase-space domain expressed by $x \in (-\infty,\,+\infty)$ and $k \in (-\infty,\,+\infty)$, in order to have
\begin{equation}
\mathcal{Z}_0(\beta)=
\int^{+\infty}_{-\infty} \hspace{-.35cm}dx\,\int^{+\infty}_{-\infty} \hspace{-.35cm}dk\,
\mathcal{W}_0(x,\,k;\,\, \beta)=
4\,K_0(\beta)\,K_0(a\,\beta),\end{equation} 	 
where $K_\nu(s)$ is the modified Bessel function of the second kind of order $\nu$~\cite{Gradshteyn}.

After some straightforward mathematical manipulations, from Eq.~\eqref{oideem} one obtains
\begin{eqnarray}\label{oideem2}
\mathcal{W}^{(2)}_{ST}(x,\,k;\,\, \beta) &=& \frac{\mathcal{Z}_0\, (\beta)}{ \mathcal{Z}_{ST}\, (\beta)}\,\mathcal{W}_{0}(x,\,k;\,\beta) \,
\left[1+\varepsilon_{(x,\,k;\,\beta)}
\right],
\end{eqnarray}	
with
\begin{equation}
\varepsilon_{(x,\,k;\,\beta)}=
\frac{a \beta^2}{8} \cosh(k)\cosh(x)\left[
\frac{\beta}{3}
\left(
a\, \tanh(x)\sinh(x) + \tanh(k)\sinh(k) 
\right)
- 1
\right],\end{equation}
such that the resulting partition function is given by
\begin{equation}\mathcal{Z}_{ST}(\beta)=
4\left(K_0(\beta)\,K_0(a\,\beta)-\frac{a\,\beta^2}{24}K_1(\beta)\,K_1(a\,\beta)\right) \, \,
\end{equation} 
and which leads to associated Wigner currents given by
\begin{eqnarray}
\mathcal{J}^{(2)}_x(x, \, k;\,\tau) &=& \sinh(k)\left\{1+\varepsilon_{(x,\,k;\,\beta)} - 
\frac{a\,\beta}{24}\left[
a\,\beta \sinh^2(x) - \cosh(x)
\right]
\right\}
\mathcal{W}_{0},\\
\mathcal{J}^{(2)}_k(x, \, k;\,\tau) &=& -a\,\sinh(x)\left\{1+\varepsilon_{(x,\,k;\,\beta)} -
\frac{\beta}{24}\left[
\beta \sinh^2(k) - \cosh(k)
\right]
\right\}
\mathcal{W}_{0},
\end{eqnarray}	
and to the Liouvillianity quantifier,
\begin{equation}\label{groms}
\mbox{\boldmath $\nabla$}_{\xi} \cdot \mathbf{w}=\frac{a\, \beta^2}{12}\sinh(x) \sinh(k) 
\left[a\,\cosh(x) -\cosh(k)\right].
\end{equation}

\begin{figure}[h]
\includegraphics[scale=0.284]{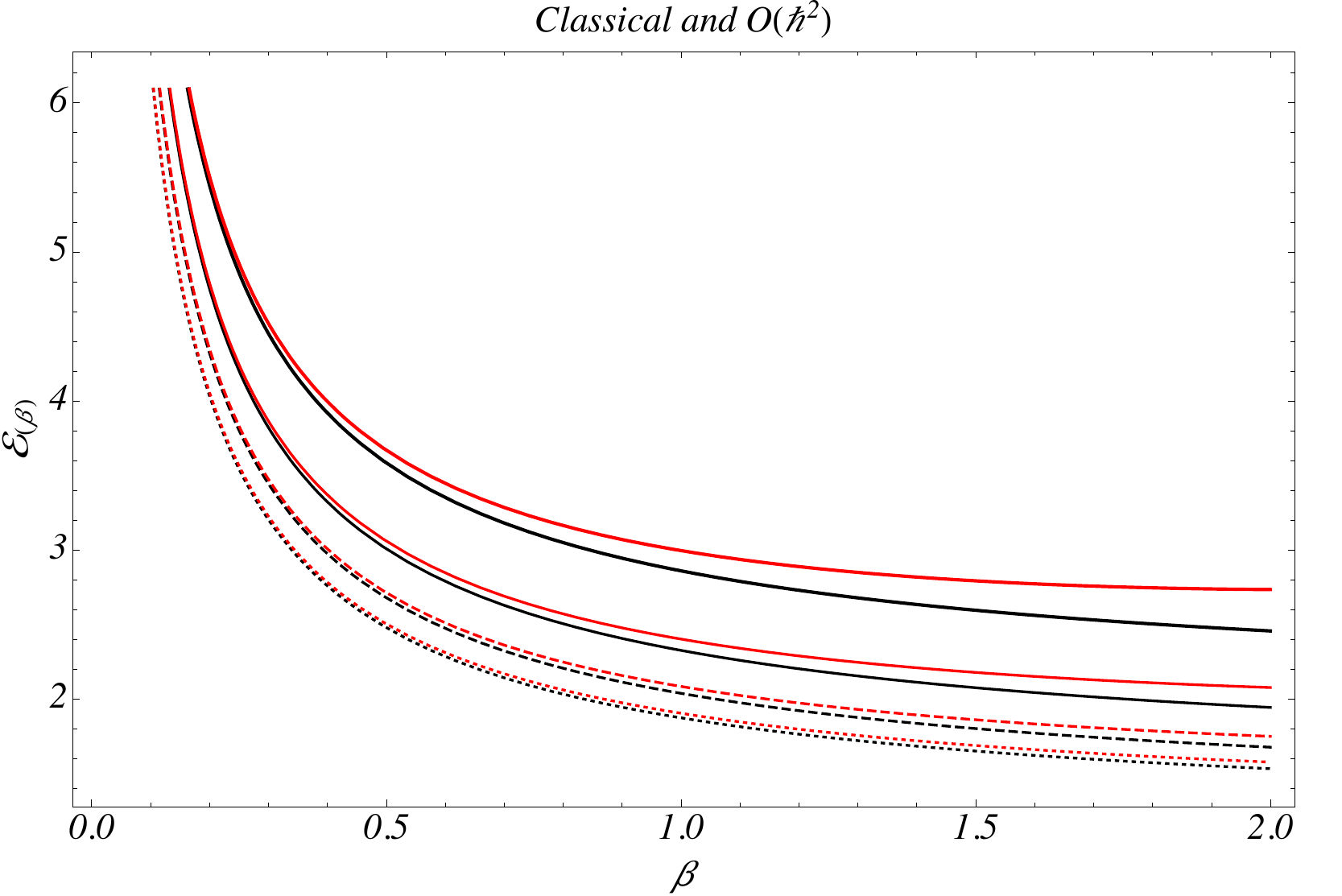}\hspace{-.1 cm}
\includegraphics[scale=0.29]{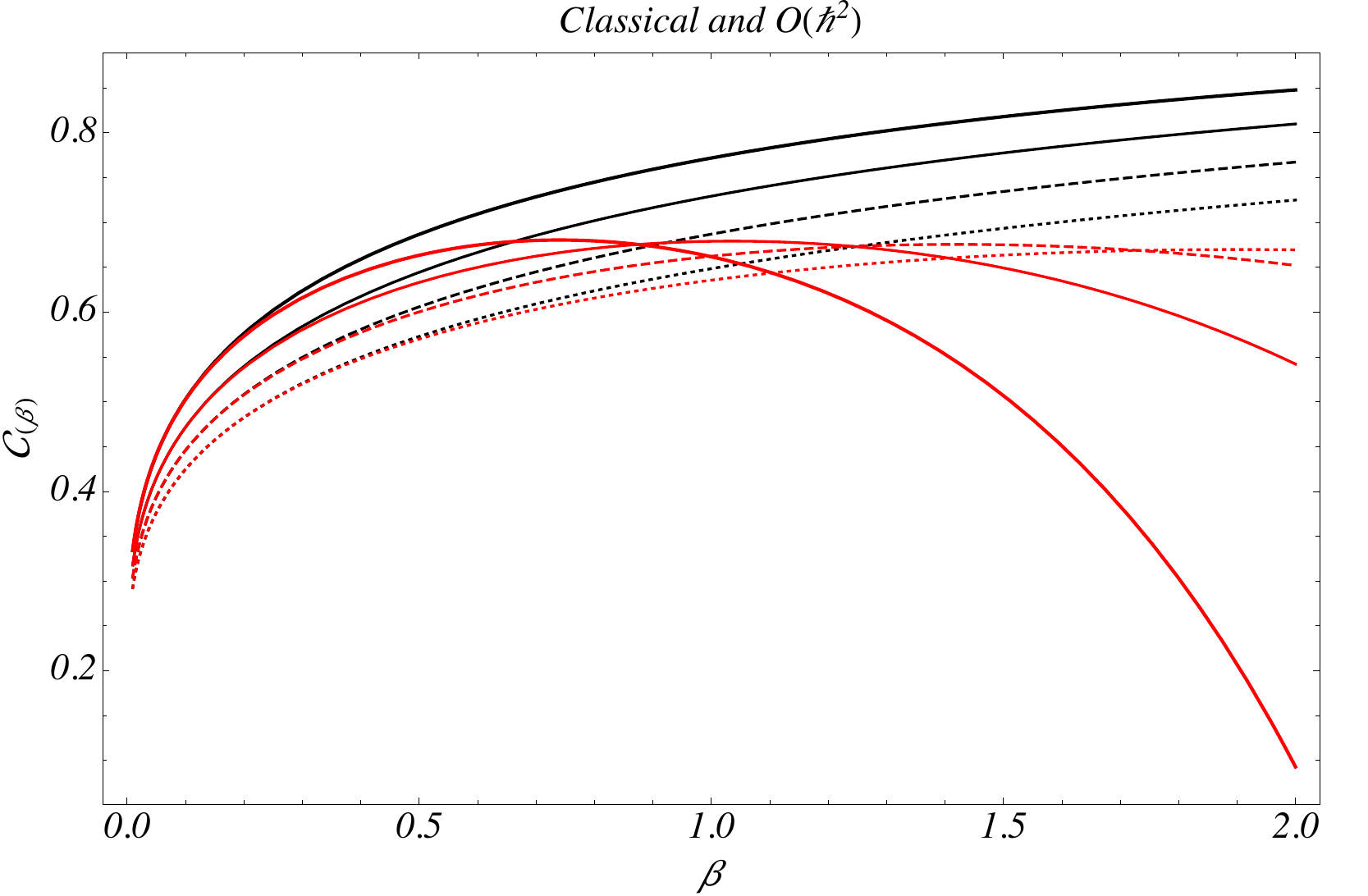}\
\renewcommand{\baselinestretch}{.6}
\caption{\footnotesize
(Color online) Internal energy (first plot), $\mathcal{E}(\beta)$, and heat capacity (second plot), $\mathcal{C}(\beta)$, as function of $\beta$, for classical (black lines) and quantum ( $\mathcal{O}(\hbar^2)$) (red lines) stationary regimes.
The results are for $a=1/2$ (dotted lines), $1$ (dashed lines), ${2}$ (thin lines) and $4$ (thick lines).}
\label{HprHpr02}
\end{figure}

Admitting that microscopic systems can be statistically described by Toda-like Hamiltonians, specially in the context of prey-predator dynamics as, for instance, if it has been identified by molecular and biochemical systems~\cite{PP00,PP01,PP02,PP03,PP04}, the quantum ($\mathcal{O}(\hbar^2)$ corrected) partition function, $\mathcal{Z}_{ST}(\beta)$, allows for obtaining the associated (dimensionless) internal energy, $\mathcal{E}_{\, \beta} (\equiv E/(\hbar\omega))$, and heat capacity, $\mathcal{C}_{\, \beta}(\equiv C/k_B)$, 
$$
\mathcal{E}_{\, \beta}=-\frac{\partial~}{\partial \, \beta}\ln\left[{\mathcal{Z}(\beta)}\right]
,\quad\mbox{and} \quad
 \mathcal{C}_{\, \beta}=\beta^2\left(\frac{\partial~}{\partial \, \beta}\right)^2\ln\left[{\mathcal{Z}(\beta)}\right],$$
which are depicted in Fig.~\ref{HprHpr02}, in comparison with classical results.
One notices that the classical pattern is just slightly affected by quantum effects, namely for higher temperatures, where the approach applies.
Of course, a phase-space quantum distorting map is just sensible to $\mathcal{O}(\hbar^2)$ corrections which are only evinced for $\beta_{\hbar \omega} \gg 1$. For such distortions to make sense as quantum fluctuations, for a $\beta$ parameter arbitrarily chosen, one should have $\beta k_B \mathcal{T}/\omega=\hbar \ll 1$ in order to guarantee the assertiveness of the $\mathcal{O}(\hbar^2)$ approximation.
Gaussian ensembles, for example, are much more sensible to higher order quantum corrections, since that, as it shall be obtained in the following section, the overall contributions due to the $\hbar^2$ series expansion results into exact expressions for the Wigner current components, providing the conditions for a non-perturbative analysis.

\subsection{Non-perturbative evaluation of Gaussian ensembles}

Once replaced into Eqs.~\eqref{imWA} and \eqref{imWB}, the dimensionless form of Gaussian Wigner functions\footnote{Considering position and momentum variables, $q$ and $p$, one should have
\begin{equation}
G_\alpha(q,\,p)=\frac{\alpha^2}{\pi\hbar}\, \exp\left[-\frac{\alpha^2}{\hbar}\left(\frac{q^2}{\mathcal{A}^2}+ \mathcal{A}^2\,p^2\right)\right],
\end{equation}
as to obtain $\mathcal{G}_\alpha(x,\,k)=\hbar \,G_\alpha(q,\,p) $ where the parameter $a$ is identified by $\mathcal{A}=(m\,\omega)^{-1}$, the mass scale is identified by $m$, for the same already introduced (cf. the Appendix) arbitrary angular frequency, $\omega$.} expressed by
\begin{equation}\label{gas}
\mathcal{W}(x, \, k;\,\tau) \equiv \mathcal{W}(x(\tau), \, k(\tau);\,0)\to\mathcal{G}_\alpha(x,\,k)=\frac{\alpha^2}{\pi}\, \exp\left[-\alpha^2\left(x^2+ k^2\right)\right],
\end{equation}
shall drive its associated fluid-analog pattern through the Wigner current components written as
\begin{eqnarray}
\label{imWA22}\partial_x\mathcal{J}^{\alpha}_x(x, \, k;\,\tau) &=& +\sum_{\eta=0}^{\infty} \left(\frac{i}{2}\right)^{2\eta}\frac{1}{(2\eta+1)!} \, \left[\partial_k^{2\eta+1}\mathcal{K}(k)\right]\,\partial_x^{2\eta+1}\mathcal{G}_{\alpha}(x, \, k),
\\
\label{imWB22}\partial_k\mathcal{J}^{\alpha}_k(x, \, k;\,\tau) &=& -\sum_{\eta=0}^{\infty} \left(\frac{i}{2}\right)^{2\eta}\frac{1}{(2\eta+1)!} \, \left[\partial_x^{2\eta+1}\mathcal{V}(x)\right]\,\partial_k^{2\eta+1}\mathcal{G}_{\alpha}(x, \, k).
\end{eqnarray}
Considering the Gaussian derivative properties given by
\begin{equation}\label{ssae}
\partial_\chi^{2\eta+1}\mathcal{G}_{\alpha}(x, \, k)=(-1)^{2\eta+1}\alpha^{2\eta+1}\,\mbox{\sc{h}}_{2\eta+1} (\alpha \chi)\, \mathcal{G}_{\alpha}(x, \, k),\qquad\mbox{for $\chi=x,\, k$,}
\end{equation}
where $\mbox{\sc{h}}_n(\alpha \chi)$ are the Hermite polynomials of order $n$, the series expansions from Eqs.~\eqref{imWA22} and \eqref{imWB22} result into convergent expressions~\cite{Novo2022,Novo2022B} which in this case encompass all the quantum contributions to the Wigner flow in a non-perturbative final form.

By replacing the kinetic and potential contributions, $\mathcal{K}(k)$ and $\mathcal{V}(x)$, from Toda-like Hamiltonian Eq.~\eqref{Original} into Eqs.~\eqref{imWA22} and \eqref{imWB22}, one identifies the derivatives given by
\begin{eqnarray}
\label{t111B}
\partial_x^{2\eta+1}\mathcal{K}(k) &=& \sinh(k),\\
\label{t222B}
\partial_k^{2\eta+1}\mathcal{V}(x) &=& a \,\sinh(x), 
\end{eqnarray}
which, after noticing that the Hermite polynomial property,
\begin{equation}
\sum_{\eta=0}^{\infty}\mbox{\sc{h}}_{2\eta+1} (\alpha \chi)\frac{s^{2\eta+1}}{(2\eta+1)!}=\sinh(2s\,\alpha\chi) \exp[-s^2],
\end{equation}
yields\footnote{A subtle feature depicted by the above analysis can be noticed by setting $\alpha^2=i$ so to have the Wigner function
\begin{equation}
\mathcal{G}_{\sqrt{i}}(x,\,k)=i\pi^{-1} \, \exp\left[- i\left(x^2+ k^2\right)\right],
\end{equation}
such that Eqs.~\eqref{imWA4CC3} and \eqref{imWB4CC3} can be recast as
\begin{eqnarray}
\label{imWA4CC322}
\partial_x\mathcal{J}^{\alpha}_x(x, \, k;\,\tau) &=& \frac{2}{\pi}e^{\frac{i}{4}} \sinh(k)\,\sinh\left(x\right)\,\exp\left[-i \left(x^2+ k^2\right)\right],\\
\label{imWB4CC322}
\partial_k\mathcal{J}^{\alpha}_k(x, \, k;\,\tau) &=& -\frac{2a}{\pi}e^{\frac{i}{4}} \sinh(k)\,\sinh\left(x\right)\,\exp\left[-i \left(x^2+ k^2\right)\right],
\end{eqnarray}
which, for $a=1$, leads to $\mbox{\boldmath $\nabla$}_{\xi} \cdot \mbox{\boldmath $\mathcal{J}^{\alpha}$}=0$, i.e. the stationary behavior of the Wigner function identified by $\mathcal{G}_{\sqrt{i}}(x,\,k)$.}
\begin{eqnarray}
\label{imWA4CC3}
\partial_x\mathcal{J}^{\alpha}_x(x, \, k;\,\tau) &=& 2 \sinh(k)\,\sin\left(\alpha^2\,x\right)\,e^{\frac{\alpha^2}{4}}\,\mathcal{G}_{\alpha}(x, \, k),\\
\label{imWB4CC3}
\partial_k\mathcal{J}^{\alpha}_k(x, \, k;\,\tau) &=& -2 a\,\sinh(x)\sin\left(\alpha^2\,k\right)\,e^{\frac{\alpha^2}{4}}
\,\mathcal{G}_{\alpha}(x, \, k),
\end{eqnarray}
and lead to an exact non-perturbative dependence on the Planck's constant, $\hbar$, implicitly yielded in terms of $x=\left(m\,\omega\,\hbar^{-1}\right)^{1/2} q$ and $k=\left(m\,\omega\,\hbar\right)^{-1/2}p$.

From Eqs.~\eqref{imWA4CC3} and \eqref{imWB4CC3}, position and momentum integrations lead respectively to the corresponding Wigner currents
\begin{eqnarray}
\label{imWA4CCD4}\mathcal{J}^{\alpha}_x(x, \, k;\,\tau) &=& 
-\frac{i\,\alpha}{2\sqrt{\pi}} \,\sinh(k)\,e^{-\alpha^2 k^2}
\left\{\mbox{\sc{Erf}}\left[\alpha(x-i/2)\right]-\mbox{\sc{Erf}}\left[\alpha(x+i/2)\right]\right\},\\
\label{imWB4CCD4}\mathcal{J}^{\alpha}_k(x, \, k;\,\tau) &=& 
+\frac{i\,a\,\alpha}{2\sqrt{\pi}} \,\sinh(x)\,e^{-\alpha^2 x^2}
\left\{\mbox{\sc{Erf}}\left[\alpha(k-i/2)\right]-\mbox{\sc{Erf}}\left[\alpha(k+i/2)\right]\right\},\,\,\,\,
\end{eqnarray}
where $\mbox{\sc{Erf}}[\dots]$ corresponds to the Gaussian error function.

\begin{figure}
\vspace{-1.4 cm}\includegraphics[scale=0.2]{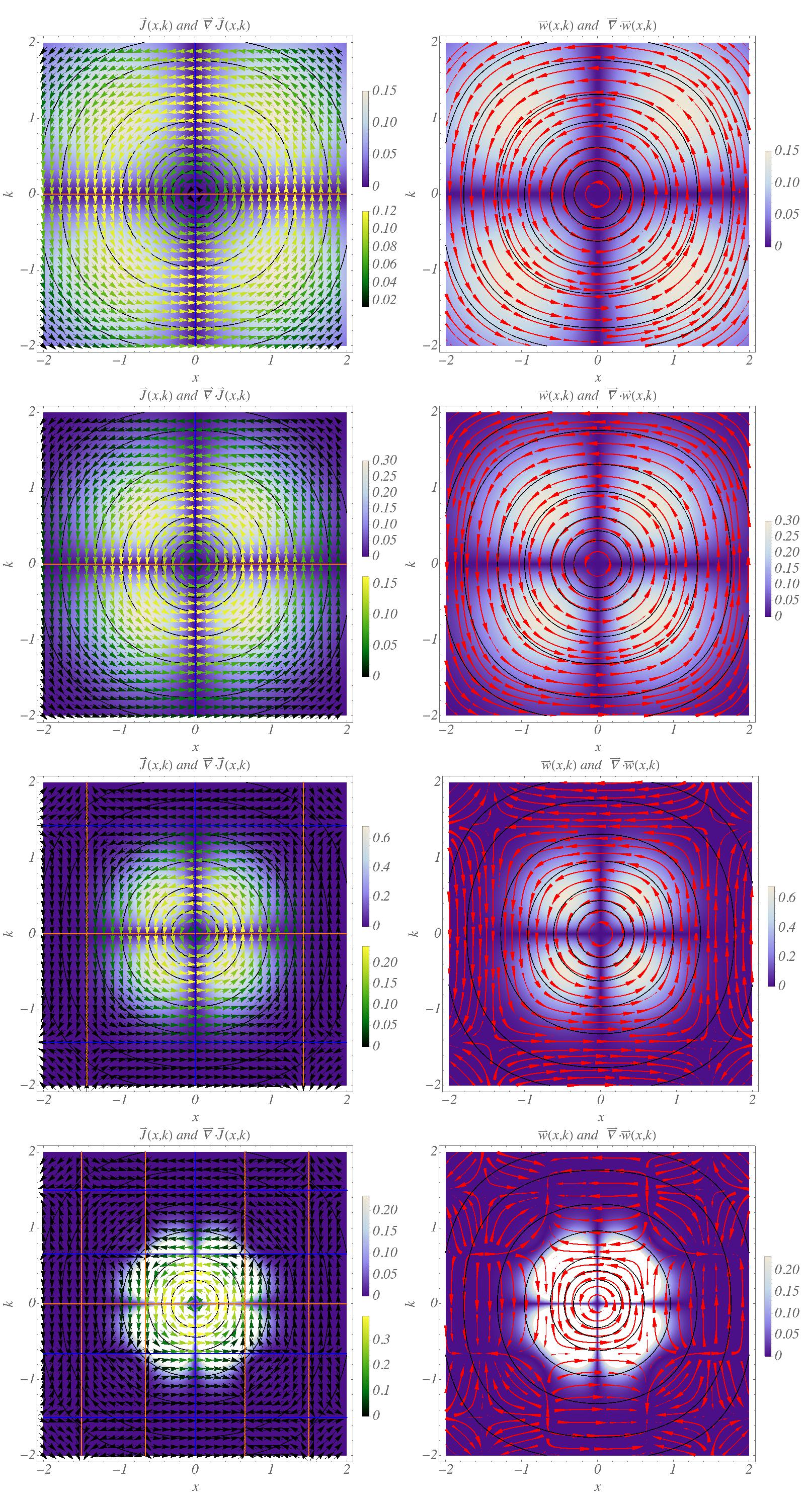}
\renewcommand{\baselinestretch}{.6}
\vspace{-.6 cm}\caption{\footnotesize
(Color online)
{\em First column}: Phase-space picture of the Wigner currents, in a vector scheme, and of the stationarity quantifier, $\mbox{\boldmath $\nabla$}_{\xi} \cdot \mbox{\boldmath $\mathcal{J}^{\alpha}$}$, in a background color scheme, for Gaussian ensembles. The magnitude of $\mbox{\boldmath $\mathcal{J}^{\alpha}$}$ is given by the {\em greenyellow} color scheme.
The results are for the increasing values of $\alpha$, from $\alpha =1/\sqrt{2}$ (first row), $1$ (second row) and $\sqrt{2}$ (third row). Peaked Gaussians ($\alpha =\sqrt{2}$) localize and discretize the quantum distortions by delimiting well-defined domain walls, which a centered non-stationarity and non-liovillianity pattern. Spread Gaussians ($\alpha =1/\sqrt{2}$), in some sense, destroy the quantum pattern as they approach the classical limit.
{\em Second column}: The normalized velocities (red arrows), $\mathbf{w}$, and the Liouvillian quantifier (background color scheme), $\mbox{\boldmath $\nabla$}_{\xi} \cdot \mathbf{w}$. The Liouvillian pattern runs from darker regions, $\mbox{\boldmath $\nabla$}_{\xi} \cdot \mathbf{w} \sim 0$, to lighter regions, $\mbox{\boldmath $\nabla$}_{\xi} \cdot \mathbf{w} > 0$. 
The classical pattern is shown as a collection of black lines. The quantum results are just for the departing configurations at $\tau=0$.}
\label{Lotka04-MD}
\end{figure}

Exploring the Gaussian ensemble properties, one also notices that the parametric constraint given by $\mbox{\boldmath $\mathcal{J}^{\alpha}$}=\mathbf{w}\,\mathcal{G}_{\alpha}(x, \, k)$, where the quantum-like velocity, $\mathbf{w}$, is identified in terms of $w_x$ and $w_k$ components, bring up some suitable topological properties of the Wigner flow, which can be depicted in the first column from Fig.~\ref{Lotka04-MD}.

The Liouvillian quantifier, $\mbox{\boldmath $\nabla$}_{\xi} \cdot \mathbf{w}$, and the quantum-like velocity, $\mathbf{w}$, are depicted by the second column from Fig.~\ref{Lotka04-MD}, where it is compared with the classical pattern identified as a collection of black thin lines. 
The quantum distortions are identified by the Wigner current properties which emerge as typical patterns from quantum effects.
The Wigner flow stagnation points are identified by orange-blue crossing lines, where $\mathcal{J}^{(2)}_x = \mathcal{J}^{(2)}_k=0$, and can be interpreted as a consequence of the $\mathcal{O}(\hbar^2)$ and higher order (quantum) contributions.
Blue (orange) contour lines are bound for the reversal of the Wigner flow in the $x(k)$ direction.
Separatrix intersections and saddle points, both with normalized circulation numbers equal to $0$, are identified, as well as clockwise and anti-clockwise vortices, with normalized circulation numbers equals to $+1$ and $-1$, respectively\footnote{The two-dimensional vorticity of a generic phase-space velocity, $\mathbf{w}$, is given by~\cite{Scripta}
\begin{equation}\label{rotat}
(\mbox{\boldmath $\nabla$}_{\xi}\times {\mathbf{w}})\cdot\hat{\mathbf{z}} = {\partial_x {\mbox{w}}_k} - {\partial_k {\mbox{w}}_x},
\end{equation}
with $\hat{\mathbf{z}}$ identified according to its orthonormality properties: $\hat{\mathbf{z}}\cdot \mathbf{v}_{\xi(\mathcal{C})} =\hat{\mathbf{z}}\cdot\mathbf{n}=0$ 
for the unitary vector $\mathbf{n} =(\dot{k},\,-\dot{x}) \vert \mathbf{v}_{\xi(\mathcal{C})}\vert^{-1} \equiv ({\partial_x \mathcal{H}},\,{\partial_k \mathcal{H}}) \vert\mathbf{v}_{\xi(\mathcal{C})}\vert^{-1}$, with $\mathbf{v}_{\xi(\mathcal{C})}\cdot\mathbf{n}=0$.
The vorticity leads to the circulation number, $\Gamma$, described as 
\begin{eqnarray}
\Gamma 
&=&
\frac{1}{2\pi\hbar}\int_{V_{\mathcal{C}}}dV\,\left({\partial_x {\mbox{w}}_k} - {\partial_k {\mbox{w}}_x}\right)
= \frac{1}{2\pi}\oint_{\mathcal{C}}d{\hat{\ell}} \,({\mbox{w}}_k,\, -{\mbox{w}}_x) \cdot\mathbf{n}
=
\frac{1}{2\pi}\oint_{\mathcal{C}}d\varphi \,\hat{\mbox{\boldmath $
\ell$}}\cdot{\mathbf{w}},
\end{eqnarray}
which accounts for the net effect of quantum affected local spinning motions of the continuum phase-space Wigner velocity distribution, where $\hat{\mbox{\boldmath $\ell$}} = \mathbf{v}_{\xi(\mathcal{C})}/|\mathbf{v}_{\xi(\mathcal{C})}|$ (with $\hat{\mbox{\boldmath $\ell$}}\cdot\mathbf{n}=0$), $d\mathtt{V} = dk\,dx$ is a dimensionless phase-space volume unity, and $d\hat{\ell} \sim d \varphi$. For unitary vectors, $\vert\mathbf{w}\vert$, one has either $\Gamma = \pm 1$ or $\Gamma=0$.
As supposed, the result remontes the Stokes theorem, through which the flow's circulation is derived from the path integral of the velocity along a closed path.
In the classical limit, where $\mathbf{w} \to \mathbf{v}_{\xi(\mathcal{C})}$, Eq.~\eqref{rotat} can be identified by the negative value of the phase-space coordinate Laplacian of the Hamiltonian, $\mathcal{H}$,
\begin{equation}
\left({\partial_x v_{k(\mathcal{C})}} - {\partial_k v_{x(\mathcal{C})}}\right) = -{\partial^2_x \mathcal{H}}-{\partial^2_k \mathcal{H}} = -\nabla^2_{\xi} \mathcal{H}.
\end{equation}}.

For such local topological patterns, the quantum contra-flux fringes bounded by green and orange lines compensate the emergence of quantum effects to sustain the dynamics driven by the Wigner continuity equation.
As expected (cf. Fig.~\ref{Lotka04-MD}), classical trajectories (black thin lines) do not exhibit such a locally compensation phenomena.

Usually, such effects affect (classical) equilibrium configurations by introducing instabilities associated to multiple focus and nodes of quantum nature (cf. Fig.~\ref{Bio02} in the following section).
However, due to the even parity of $\mathcal{V}(x)$ and $\mathcal{K}(k)$ Hamiltonian contributions, instability outputs are mutually cancelled.

The stationarity quantifier, $\mbox{\boldmath $\nabla$}_{\xi} \cdot \mbox{\boldmath $\mathcal{J}^{\alpha}$}$, is described according to the background color scheme, from lighter regions corresponding to non-vanishing local contributions to $\partial_{\tau} \mathcal{G}_{\alpha}(x,\,k)$. In particular, one notices that stationarity and Liouvillianity are smoothly decoupled one from each other. This is due to the Gaussian peaking and spreading effects which affect the Wigner currents.

As preliminarily identified from Refs.~\cite{Novo2022,Novo2022B}, Gaussian ensembles are indeed the most suitable tool for identifying semi-classical dynamical patterns which eventually map the prey-predator dynamics in the form of non-linear Hamiltonians as, for instance, that one from \eqref{Original}.
Following the conclusions of Refs.~\cite{Novo2022,Novo2022B}, the {\em quantum analog} hypothesis is subjacent to the non-commutative property, $[x,\,k] \neq 0$, which results into the phase-space local identification of quantum effects. For Gaussian ensembles, it results into an explicit reformulation of the equations of motion, with $\mathbf{v}$ replaced by $\mathbf{w}$, where quantum effects without any kind of approximative features are identified.
At least from an effective perspective, the hypothesis of quantum features affecting the dynamical evolution of some kind of ecological microscopic systems can be understood in terms of the above results.
As it shall be noticed in the next section, the framework can now fit species evolution in corresponding prey-predator systems in a way that the quantum imprints can be evinced.

\section{Semi-classical dynamical patterns for Gaussian ensembles}

\begin{figure}[h]
\includegraphics[scale=0.45]{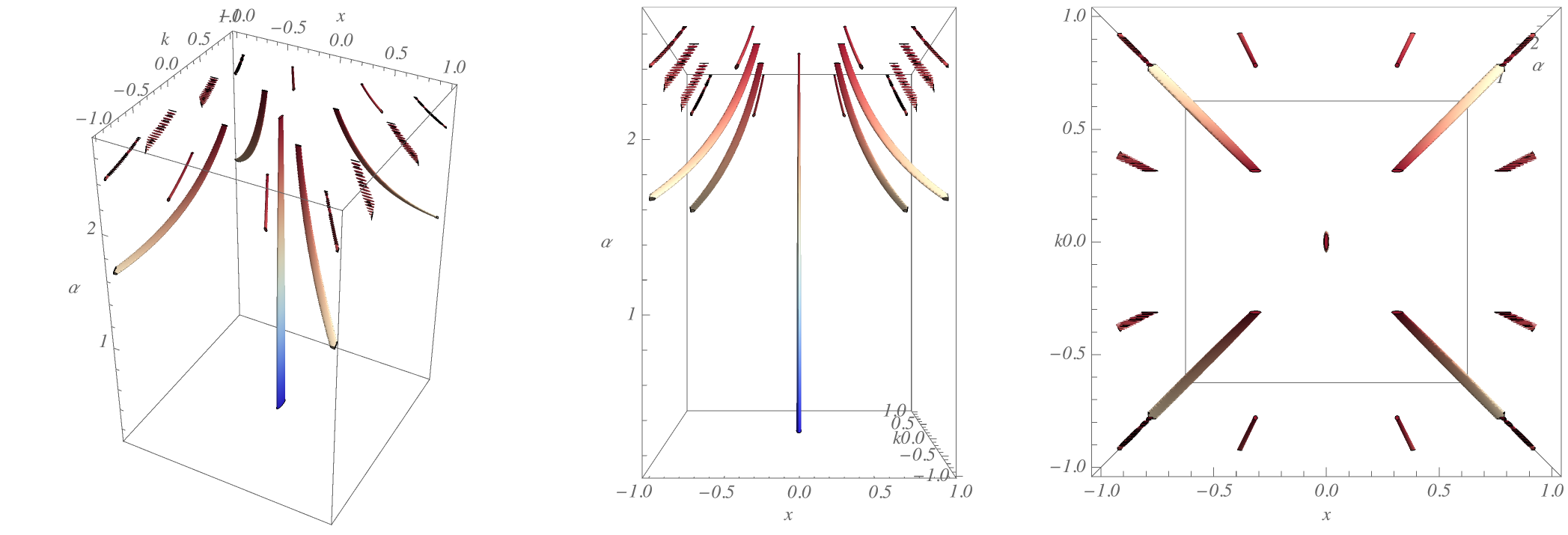}
\renewcommand{\baselinestretch}{.6}
\caption{\footnotesize
(Color online)
Region plot scheme for the phase-space evolution of quantum critical points corresponding to classical equivalent (blue regions) equilibrium points and the subsets of locally stable (white to red regions) equilibrium points driven by the Gaussian spreading parameter, $\alpha$, from $0$ (blue tones) to $2.7$ (red tones) and with
$a=4$.
Results are for flux surrounding envelops with boundaries identified by $\vert\mathbf{w}\vert < 0.08$. Quantum effects emerge thorough a compensation rate where either two vortices of opposite winding numbers match each other or saddle points that mutually annihilates one each other.
The spreading behavior of the Gaussian ensemble, from red ``bubble'' (unstable) islands to the blue ({\em quasi}-stable) envelop, corresponding to decreasing values of $\alpha$, diffusively recovers the classical-like pattern for which the effective quantum imprints are described in Fig.~~\ref{Bio03-DM} just as an equivalent prey-predator dephasing effect.\label{Bio02}}
\end{figure}

The Gaussian Wigner flow pattern obtained in the previous section can also be constrained by the form $\mbox{\boldmath $\mathcal{J}^{\alpha}$}=\mathbf{w}\,\mathcal{G}_{\alpha}(x, \, k)$, such that the equilibrium points, corresponding to $\mathcal{J}^{(2)}_x=\mathcal{J}^{(2)}_k=0$, can be evaluated in terms of the parameter $\alpha$, thus revealing a flow pattern for quantum fluctuations.
As depicted in Fig.~\ref{Bio02}, for $0 \lesssim \alpha \lesssim 2.7$, quantum-affected equilibrium points with flux-surrounding envelopes (bounded by $\vert\mathbf{w}\vert < 0.08$) are shown through multiple viewing angles.
Despite resembling classical closed orbits for $\alpha \lesssim 1.7$, these are perturbed by quantum vortex distortions that emerge from nearby values of $x$ and $k$, resembling a kind of phase transition that generates subsets of locally stable domains with independent equilibrium points.

\begin{figure}
\includegraphics[scale=0.37]{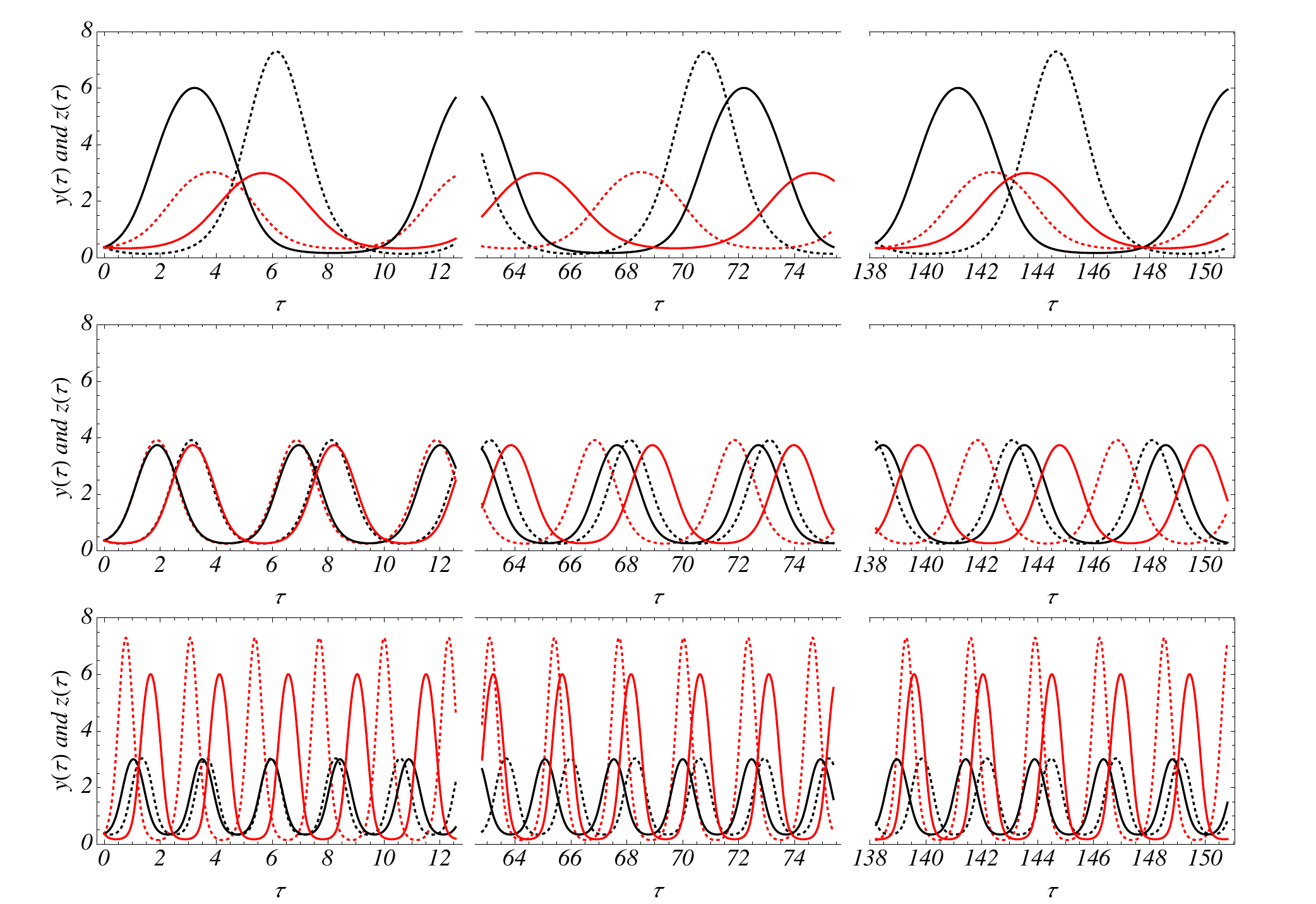}
\includegraphics[scale=0.243]{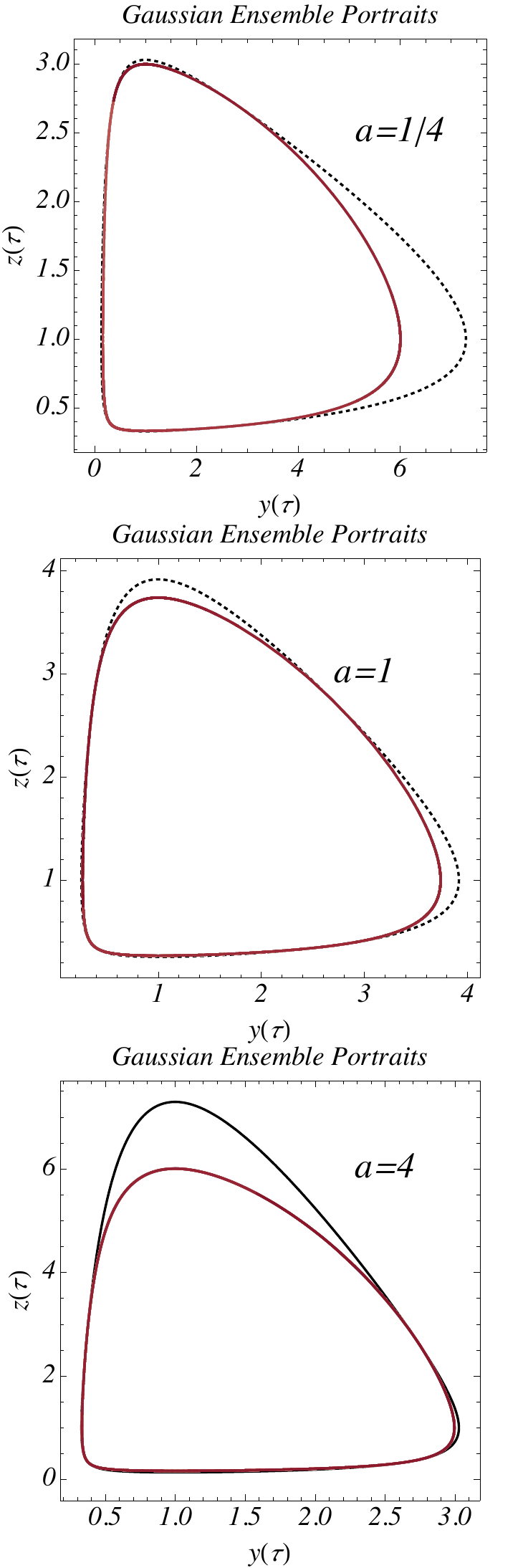}
\renewcommand{\baselinestretch}{.6}
\caption{\footnotesize
(Color online)
{\em First column}: Classical (dashed lines) and quantum (solid lines) prey-predator time-evolution pattern, $y(\tau)$ (red lines) and $z(\tau)$, for typical spreading Gaussian ensembles, with $\alpha=1$, and with $a=1/4,\,1$ and $4$.
{\em Second column}: Associated phase-space equilibrium trajectories for classical (dashed lines) and quantum (solid lines) prey-predator patterns.
.}
\label{Bio03-DM}
\end{figure}

Such a global and diffused appearance of quantum vortices and saddle points naturally disrupts the classical pattern.
Nevertheless, classical behavior persists locally, as can be more clearly observed in Fig.~\ref{Bio03-DM}. In that case, the macroscopic modifications to the equivalent prey-predator oscillation pattern induced by these topological phases can be better understood.

In fact, the quantum-modified equilibrium patterns that describe species evolution can be interpreted using the semi-classical approach to quantum-like velocities, accounting for distortions obtained from Eqs.~\eqref{imWA4CCD4} and \eqref{imWB4CCD4}.

If one assumes that the Toda-like Hamiltonian classically parameterizes a competitive dynamic where $x$ and $k$ commute (i.e., $[x,\,k]=0$), simultaneous measurements of $x$ and $k$ do not introduce any additional constraints beyond those imposed by the Hamiltonian itself.
Given that fluctuations in the number of species, $y(\tau)=e^{-x(\tau)}$ and $z(\tau)=e^{-k(\tau)}$, can be parameterized by $\delta x$ and $\delta k$, a deterministic evolution follows, as shown in Fig.~\ref{LVLVLV}. 

However, under the quantum-analog assumption of non-commutativity, $[x,\,k] \neq 0$, measurements of $x$ and $k$ affect one another, and thus cannot be performed simultaneously, i.e. $\delta x\,\delta k \neq 0$, which results in the {\em quantum analog} of the uncertainty principle, expressed in its dimensionless form as $\delta x,\delta k \gtrsim 1$, consistent with $[x,\,k]=i$.
Nonetheless, as will be shown next, this applies only to the full, unbounded phase-space domain. In fact, Eqs.~\eqref{imWA4CCD4} and \eqref{imWB4CCD4} implicitly show that deterministic semi-classical regimes can still be locally recovered.

To clarify this point, after some numerical procedure over the quantum analog phase-space velocity components, Eqs.~\eqref{imWA4CCD4} and \eqref{imWB4CCD4}, one straightforwardly obtains the dynamical behavior of equivalent prey and predator species, $y(\tau)$ and $z(\tau)$.
As depicted in Fig.~\ref{Bio03-DM}, for typical spreading Gaussian ensembles with $\alpha=1$, and with $a=1/4,\,1$ and $4$, one notices that the equilibrium point is not macroscopically affected by quantum fluctuations.
Irrespectively of $a$ values, there is just a dephasing effect between quantum and classical patterns, which can be suppressed/amplified according to the Gaussian spreading parameter, $\alpha$.
Note that the deviations provided by $a\neq 1$ just indicate positive ($a < 1$) and negative ($a > 1$) fluctuations in the prey-predator rate, $y(\tau)/z(\tau)$, without affecting the regime stability at $y=z=1$.

It is relevant to notice that once the emergence of quantum patterns and classical to quantum transitions involve some loss of stability, this could be stratified according to the hyperbolic stability criteria~\cite{Novo2022}.
For typical LV systems driven by the Hamiltonian Eq.~\eqref{Ham}, quantum corrections for anisotropic population oscillation patterns ($a\neq 0$) affect the stability pattern~\cite{Novo2022}.

For LV systems in the regime $a<0$ with $\alpha\neq 0$, quantum corrections lead, in the long-time limit, to a dynamics characterized by recurrence of population collapses followed by revivals. This behavior is closely related to earlier findings reported in Ref.~\cite{PRE-LV}, where the discrete nature of populations undermines the mean-field stability and ultimately results in the disappearance of one or both species.
By contrast, when $a>0$ and $\alpha\neq 0$, the oscillatory behavior of the populations is progressively damped, so that the dynamics tends to settle near equilibrium configurations. Notably, this outcome is consistent with the phenomenon known as {\it extinction of oscillating populations} discussed in Ref.~\cite{PRE-LV2}, and it also displays qualitative similarities with the patterns reported in Ref.~\cite{PRE-LV4}, which analyzes a stochastic formulation of the Rosenzweig-MacArthur predator-prey model~\cite{Mac63}.

Definitely, this is not the case for Toda-like Hamiltonians.

As previously mentioned, classical equilibrium configurations are not qualitatively affected by the presence of multiple focus and nodes of quantum nature (cf. Fig.~\ref{Bio02}). Due to the even parity of $\mathcal{V}(x)$ and $\mathcal{K}(k)$ Hamiltonian contributions, Eq.~\eqref{Original}, instability outputs are mutually cancelled.
The persistence of closed-orbit equilibrium dynamics is supported by the stability of the equilibrium point at $y=z=1$ ($x=k=0$), which directly follows from the equations of motion for this quantum-like Toda regime. In fact, it is identified by the defined parity pattern of the Wigner currents (as computed from Eqs.~\eqref{imWA4CCD4} and~\eqref{imWB4CCD4}) depicted in Fig.~\ref{Lotka04-MD}.
Therefore, from a quantum mechanical perspective, Toda-like patterns are significantly more stable than LV patterns~\cite{Novo2022C}.
This represents the most significant distinction compared to previous results for quantum-like LV regimes~\cite{Novo2022B}, in which, despite the identification of attractor behaviors, quantum fluctuations ultimately destroy the closed-orbit structure of the phase-space.

\section{Conclusions}

Phase-space features of the synthetic version of the Toda-like Hamiltonian, written in a form constrained by the condition $\partial^2 \mathcal{H} / \partial x \partial k=0$, were investigated in the framework of the WW QM.
TD and Gaussian quantum ensembles were analytically constructed to identify corrections arising from quantum distortions over classical phase-space patterns, quantified in terms of Liouvillian and stationarity-related differential operators.

From a non-deterministic perspective, the prediction of the existence of competitive species related to the Hamiltonian description here addressed, considering the hypothesis of interacting quantum states~\cite{Bio17}, $\hat{x}$ and $\hat{k}$ can be viewed as measurement operators. This, in turn, can be associated with a quantum statistical ensemble description, for example, through a Gaussian phase-space probability distribution. Hence, the interpretation of ecosystem components via quantum statistical distributions may follow the statistics of Gaussian ensembles\footnote{In classical and quantum frameworks.}.

This implies that collective behaviors quantified by statistics outcomes derived from phase-space semiclassical trajectories could represent the averaged results from the statistical treatment of the space-time evolution of the species distribution densities, $y$ and $z$: if quantum observables are defined by canonically conjugate operators, $\hat{x}$ and $\hat{k}$, such that their averaged values, $x = \langle \hat{x} \rangle$ and $k = \langle \hat{k} \rangle$, are related to the species distribution densities via $y = e^{-x}$ and $z = e^{-k}$, a measurement operation is affected by the fundamental quantum-like non-commutative property of the phase-space coordinates, $\hat{x}$ and $\hat{k}$, $[\hat{x}, \hat{k}] \neq 0$. 

Indeed, the information incompleteness tied to the probabilistic nature of the quantum-like non-commutative assumption, $[\hat{x}, \hat{k}] \neq 0$, necessitates a complementary statistical view of the dynamically involved variables, $x$ and $k$.
Specifically, if $[\hat{x}, \hat{k}] = i$, where the Planck constant, $\hbar$, is set to unity, both classical and quantum dynamics can be assumed to coexist at different scales: classical at macroscopic levels and quantum at microscopic levels.

In the classical regime, described by the series expansions in Eqs.~\eqref{imWA} and \eqref{imWB} truncated at $\eta = 0$, the same result for $x$ and $k$ satisfying $[\hat{x}, \hat{k}] = 0$ is recovered. In this case, species number fluctuations, represented by second-order momenta $\delta s = \sqrt{\langle \hat{s}^2 \rangle - \langle \hat{s} \rangle^2}$, evolve deterministically, with $\delta x \, \delta k = 0$, governed by the classical velocity, $\mathbf{v}_{\xi(\mathcal{C})}$.
The {\em quantum analog} introduces a non-commutative constraint, $[\hat{x}, \hat{k}] = i$, implying a minimal phase-space volume $\delta x \, \delta k \sim 1$. This defines a {\em quantum-origin} non-extinction hypothesis, where measurements of $\hat{x}$ and $\hat{k}$ (and the corresponding species densities $y$ and $z$) cannot be simultaneously made. This results in the {\em quantum analog} of the uncertainty principle: $\delta x \, \delta k \gtrsim 1$. Consequently, species evolution becomes non-deterministic, and the fluctuations in species numbers cannot be parametrized by $y(\tau)$ and $z(\tau)$ or $y(z) \leftrightarrow z(y)$. Wigner currents then drive the statistical effects within quantum ensembles, generating semiclassical trajectories~\cite{Novo2022B}.

For quantized Toda-like model, the fluctuations in species number, represented by $\delta x$ and $\delta k$, exhibit a deterministic pattern derived from a periodic evolution parameterized by Wigner currents. Despite exhibiting a non-Liouvillian behavior, with $\mbox{\boldmath $\nabla$}_{\xi} \cdot \mathbf{w} \neq 0$, the system admits a {\em quantum analog} semi-classical regime, in which closed phase-space orbits can still be identified.
When comparing our results to broader descriptions of quantum-like effects in competitive biosystems, the persistence of closed-orbit equilibrium dynamics in the quantum-like Toda regime suggests that, from a quantum mechanical perspective, Toda-like patterns are significantly more stable than those of the LV model~\cite{Novo2022C}.

Only during critical transitions from classical to quantum regimes, marked by the presence of topological effects over the phase-space quantum profile, classical periodicity is quantum mechanically destroyed.
These topological drivers are locally quantified using the previously mentioned stability and Liouvillianity measures, and can be macroscopically interpreted as sudden and unexpected changes in the classical solutions of the associated anharmonic system.

From this perspective, a detailed examination of phase-space dynamics, including a systematic assessment of quantum corrections and their role in classical-quantum emergence, provides additional support for our treatment of more realistic nonlinear models. Our deterministic analysis represents a first, relevant step toward a more realistic stochastic modeling framework.

At first glance, since hyperbolic behavior plays a central role in characterizing competitive ecological equilibria, the Hamiltonian formulation of prey-predator-like systems~\cite{LV1,LV2,Novo2022C,Novo2024} -- particularly when employed in the analysis of stochastic dynamics~\cite{Allen,Grasman} -- leads to equations of motion that naturally emerge from a phase-space description. Within this setting, the associated trajectories can be consistently extended to a non-commutative framework, $[x,,k]=i$, by means of the WW formalism.

However, it must be observed that our analysis describes a closed ecosystem with no interaction with a noisy environment. In realistic biological systems, environmental fluctuations must be addressed. Indeed, previous studies have shown that noise can have nontrivial and even constructive effects in ecological dynamics: for example, spatiotemporal patterns in noisy LV systems exhibit nonmonotonic behaviors as a function of noise intensity~\cite{Valenti2004}, and stochastic models of phytoplankton dynamics highlight the importance of environmental variability in shaping population distributions in natural ecosystems~\cite{Valenti2016}.

Such models may also incorporate scenarios influenced by additional statistical constraints.
One example is the adoption of gamma distributions to describe processes composed of multiple sub-events, such as sequences of cell-division events~\cite{26} or compensatory mutations associated with a given genetic alteration~\cite{27}. Clearly, when viewed through a quantum-mechanical lens, phenomena like cell division and mutation involve extremely complex mechanisms, and one should not expect simple outcomes from direct quantum calculations~\cite{Bord13,Bord19}. Nevertheless, suitably chosen approximations can still capture essential aspects of microscopic biological systems and allow the incorporation of characteristic elements of a quantum-theoretical description~\cite{Bord13}.

As an illustrative case, quantum-mechanically induced directed adaptive mutations have been examined in the context of carcinogenesis~\cite{Bord19}. Closely related to Wigner's seminal ideas on quantum corrections to TD ensembles~\cite{Wigner,Novo2022A}, the evaluation of tunneling rates associated with spontaneous point mutations in DNA has also been investigated~\cite{DNA1}. In this context, an open quantum system approach to proton transfer in DNA~\cite{DNA2}, based on the phase-space formulation of the Caldeira-Leggett master equation~\cite{Caldeira}, leads to the Wigner-Moyal-Caldeira-Leggett equation (see Eq.~(1) of Ref.~\cite{DNA2}), which provides the theoretical framework for such calculations. The Wigner formalism thus acts as a possible bridge between microscopic biochemical systems and quantum mechanics~\cite{Novo2024}.

Moreover, the present study can also be framed within the broader context of noise-induced phenomena in complex systems. Fluctuations, rather than merely representing sources of disorder, can play constructive roles in stabilizing multiple equilibria, enhancing transport processes, and inducing transitions that are otherwise forbidden in deterministic dynamics. For instance, Parisi's review~\cite{Parisi2023} highlights the universal relevance of fluctuations across scales, from microscopic constituents to macroscopic complex systems, and emphasizes how stochastic effects can generate multiple metastable states in disordered and glassy systems. In quantum optics and light-matter systems, stochastic driving has been shown to induce correlations and cooperative behaviors~\cite{Stassi2015}, while quantum dissipative models illustrate how non-Gaussian noise can enhance coherence and signal propagation in open quantum systems~\cite{Carollo2018,Magazzu2015}. On a classical level, interacting random walkers and nonlinear networked systems display noise-induced transitions that underlie anomalous diffusion, synchronization, and emergent collective phenomena~\cite{Spezia2008,Agudov2021}. In biological, ecological, and memristive contexts, such stochastic effects underpin adaptive responses, signal amplification, and anomalous transport. By situating our LV- and Toda-like quantum-phase-space models within this conceptual landscape, one can appreciate how microscopic noise and quantum fluctuations contribute to emergent macroscopic patterns, suggesting a natural explanation for the interplay between stochasticity and structural organization in complex systems.

Finally, similar in scope, recent developments in quantum metrology and criticality highlight the interdisciplinary relevance of fluctuation-driven dynamics. Studies in statistical physics have demonstrated that near critical points, sensitivity to external perturbations is greatly amplified, allowing stochastic fluctuations to be exploited for enhanced precision and information extraction~\cite{JSTAT2009,DiFresco2022}. For example, quantum Fisher information can diverge at criticality, and even small environmental noise can influence measurement outcomes, illustrating the subtle but constructive role of fluctuations in parameter estimation. These results underscore the broader lesson that fluctuations are not merely a source of uncertainty; they can be a resource, especially when systems operate near phase transitions or critical points. In this sense, the theoretical framework employed here -- combining Hamiltonian models, stochastic dynamics, and phase-space quantum descriptions -- again suggests an interplay between microscopic noise phenomena and experimentally accessible macroscopic observables. 

Of course, the above analysis clarifies the scope and limitations of the present approach and emphasizes its possible implications for complex, living, and engineered systems. Although our results do not provide the definitive set of algorithms for evaluating quantum effects in nonlinear systems, nor to establish the ultimate answer for biological phenomenology questions, they represent a meaningful step toward these broader objectives.

\vspace{.5 cm}
{\em Acknowledgments -- This work is supported by the Brazilian Agency CNPq (Grant No. 301485/2022-4).}

\section*{Appendix -- Wigner function elementary properties} 

Defined through the Weyl transform of an operator $\hat{O}(\hat{q}, \hat{p})$,
\begin{equation}
O^W(q,\, p)\label{111}
= 2\hspace{-.1cm} \int^{+\infty}_{-\infty} \hspace{-.35cm}ds\,\exp{\left[2\,i \,p\, s/\hbar\right]}\,\langle q - s | \hat{O} | q + s \rangle=2\hspace{-.1cm} \int^{+\infty}_{-\infty} \hspace{-.35cm} dr \,\exp{\left[-2\, i \,q\, r/\hbar\right]}\,\langle p - r | \hat{O} | p + r\rangle,
\end{equation} 
where, correspondently, position and momentum operators, $\hat{q}$ and $\hat{p}$, are converted into $c$-numbers, $q$ and $p$, the Wigner function is straightforwardly identified with the density matrix operator, $\hat{\rho}=|\psi \rangle \langle \psi |$, through the overlap integral given by
\begin{equation}
 (2\pi \hbar)^{-1} \hat{\rho} \to W(q,\, p)=(\pi\hbar)^{-1} 
\int^{+\infty}_{-\infty} \hspace{-.35cm}ds\,\exp{\left[2\, i \, p \,s/\hbar\right]}\,
\psi(q - s)\,\psi^{\ast}(q + s),\label{222}
\end{equation}
which can also be identified as the Fourier transform of the off-diagonal elements of $\hat{\rho}$.
Even when considered as for accounting for quantum corrections to TD equilibrium states~\cite{Wigner}, akin to the formalism of statistical mechanics, the WW phase-space formulation encompasses the essential QM paradigms through the information content of the quasiprobability distribution function from Eq.~\eqref{222}, from which its marginal distributions obtained through integrations over the position and momentum coordinates are respectively given by
\begin{equation}
\vert \psi_q(q)\vert^2=\int^{+\infty}_{-\infty} \hspace{-.35cm}dp\,W(q,\, p)
\qquad
\leftrightarrow
\qquad
\vert \psi_p(p)\vert^2=\int^{+\infty}_{-\infty} \hspace{-.35cm}dq\,W(q,\, p),
\end{equation}
with
\begin{equation}
\psi_q(q) =
(2\pi\hbar)^{-1/2}\int^{+\infty}_{-\infty} \hspace{-.35cm} dp\,\exp{\left[-i \, p \,q/\hbar\right]}\,
\psi_p(p).
\end{equation}
Regarding the most elementary property of the Weyl transform, the trace of the product of two operators, $\hat{O}_1$ and $\hat{O}_2$, evaluated through an overlap integral over the infinite volume described by the phase-space coordinates, $q$ and $p$, as~\cite{Wigner,Case} 
\begin{equation}
Tr_{\{q,p\}}\left[\hat{O}_1\hat{O}_2\right]=h^{-1} 
\int \hspace{-.15cm}\int \hspace{-.15cm} dq\,dp \,O^W_1(q,\, p)\,O^W_2(q,\, p),
\end{equation}
allows for computing the averaged values of quantum observables, $\hat{O}$, in terms of the Weyl products in the form of
\begin{equation}
 \langle O \rangle=
\int^{+\infty}_{-\infty} \hspace{-.35cm}dp\int^{+\infty}_{-\infty} \hspace{-.35cm} {dq}\,\,W(q,\, p)\,{O^W}(q,\, p), \label{eqfive}
\end{equation}
which reflects the trace probability properties involving $\hat{\rho}$ and $\hat{O}$, $Tr_{\{q,p\}}\left[\hat{\rho}\hat{O}\right]$. In fact, it reflects a consistent QM paradigmatic interpretation constrained by the normalization condition over $\hat{\rho}$, as $Tr_{\{q,p\}}[\hat{\rho}]=1$\footnote{Such a definition also admits an statistical QM extension, from pure states to statistical mixtures, with the quantum purity expressed in terms of the trace operation, 
\begin{equation}
Tr_{\{q,p\}}[\hat{\rho}^2]=2\pi\hbar\int^{+\infty}_{-\infty}\hspace{-.35cm}dp\int^{+\infty}_{-\infty} \hspace{-.35cm} {dq}\,\,\,W(q,\, p)^2,
\label{eqpureza}
\end{equation}
built through the replacement of ${O^W}(q,\, p)$ by $W(q,\, p)$ into Eq.~\eqref{eqfive}, and which exhibits the pure state constraint, $Tr_{\{q,p\}}[\hat{\rho}^2]=Tr_{\{q,p\}}[\hat{\rho}]=1$.}.


\begin{thebibliography}{99}

\bibitem{NossoPaper}
A. E. Bernardini and O. Bertolami, {\em Non-classicality from the phase-space flow analysis of the Weyl-Wigner quantum mechanic}, EPL {\bf 120}, 20002 (2017).
\bibitem{NossoPaper19}
A. E. Bernardini and O. Bertolami, Journal of Physics: Conf. Series {\bf 1275}, 012032 (2019).
\bibitem{Novo2022}
A. E. Bernardini and O. Bertolami, {\em Noncommutative phase-space Lotka-Volterra dynamics: The quantum analog}, Phys. Rev. E {\bf 106}, 024202 (2022).
\bibitem{Novo2022B}
A. E. Bernardini and O. Bertolami, {\em Quantum Prey-Predator Dynamics: A Gaussian Ensemble Analysis}, Found. of Phys. {\bf 53}, 63 (2023).
\bibitem{Nature01}
M. Kumar, B. Ji, K. Zengler and J. Nielsen, {\em Modelling approaches for studying the microbiome}, Nat. Microbiol. {\bf 4}, 1253 (2019).
\bibitem{Nature02}
S. Butler and J. P. O'Dwyer, {\em Stability criteria for complex microbial communities}, Nat. Commun. {\bf 9}, 2970 (2018).
\bibitem{PRE-LV3}
G. Szabo and T. Czaran, {\em Phase transition in a spatial Lotka-Volterra model}, Phys. Rev. E {\bf 63}, 061904 (2001).
\bibitem{SciRep02}
T. Tahara {\em et al.}, {\em Asymptotic stability of a modified Lotka-Volterra model with small immigrations}, Sci. Rep. {\bf 8}, 7029 (2018).
\bibitem{PRE-LV2}
N. R. Smith and B. Meerson, {\em Extinction of oscillation populations}, Phys. Rev E {\bf 93}, 032109 (2016).
\bibitem{Agui01}
M. A. M. de Aguiar, E. M. Rauch and Y. Bar-Yam, {\em Invasion and Extinction in the Mean Field Approximation for a Spatial Host-Pathogen Model}, Journal of Statistical Physics {\bf 114}, 1417 (2004).
\bibitem{PRE-LV}
M. Parker and A. Kamenev, {\em Extinction in the Lotka-Volterra model}, Phys. Rev. E {\bf 80}, 021129 (2009).
\bibitem{Novo2022A}
A. E. Bernardini and O. Bertolami, {\em Generalized phase-space description of nonlinear Hamiltonian systems and Harper-like dynamics}, Phys. Rev. A {\bf 105}, 032207 (2022).
\bibitem{Novo2025}
A. E. Bernardini and O. Bertolami, {\em Phase-space quantum distorted stability pattern for Aubry-Andr\'e-Harper dynamics},
Physica D {\bf 477}, 134700 (2025).
\bibitem{Novo2024}
A. E. Bernardini and O. Bertolami, {\em Extended Weyl-Wigner phase-space framework for nonlinear systems: Typical and modified prey-predator-like dynamics}, Phys. Rev. E {\bf 110}, 034218 (2024).

\bibitem{Wigner}
E. Wigner, {\em On the quantum correction for thermodynamic equilibrium}, Phys. Rev. {\bf 40} 749 (1932).
\bibitem{Ballentine}
L. E. Ballentine, {\em Quantum Mechanics: a Modern Development}, pp. 633 (World Scientific, Singapore 1998).
\bibitem{Case}
W. B. Case, {\em Wigner functions and Weyl transforms for pedestrians}, Am. J. Phys. {\bf 76}, 937 (2008).
\bibitem{Hillery} 
M. Hillery, R. O'Connell, M. Scully and E. Wigner, {\em Distribution functions in physics: Fundamentals}, Phys. Rep. {\bf 106}, 121 (1984).
\bibitem{Zurek02}
W. H. Zurek, {\em Decoherence and the transition from quantum to classical -- Revisited}, Phys. Today {\bf 44}, 36 (1991).
\bibitem{Steuernagel3}
O. Steuernagel, D. Kakofengitis and G. Ritter, {\em Wigner flow reveals topological order in quantum phase space dynamics}, Phys. Rev. Lett. {\bf 110}, 030401 (2013).


\bibitem{PP00}
D. P. Paula {\it et al.}, {\em Detection and decay rates of prey and prey symbionts in the gut of a predator through metagenomics}, Molecular Ecology Resources {\bf 15}, 880 (2015).
\bibitem{PP01}
T. Fujii and T. Rondelez, {\em Predator-Prey Molecular Ecosystems}, ACS Nano {\bf 7}, 27 (2013).
\bibitem{PP02}
I. R. Epstein and J. A. Pojman, {\em An Introduction to Nonlinear Chemical Dynamics} (Oxford University Press, New York 1998).
\bibitem{PP03}
J. Ackermann, B. Wlotzka and J. S. McCaskill, {\em In Vitro DNA-Based Predator-Prey System with Oscillatory Kinetics},
Bull. Math. Biol. {\bf 60}, 329 (1998).
\bibitem{PP04}
B. Wlotzka and J. S. McCaskill, {\em A Molecular Predator and Its Prey: Coupled Isothermal Amplification of Nucleic Acids}, Chem. Biol. {\bf 4}, 25 (1997).
\bibitem{RPSA-LV}
Yi-An Ma and Hong Qian, {\em A thermodynamic theory of ecology: Helmholtz theorem for Lotka-Volterra equation, extended conservation law, and stochastic predator-prey dynamics}, Proceedings of the Royal Society A {\bf 471}, 20150456 (2015).
\bibitem{Allen}
L. J. S. Allen, {\em An introduction to stochastic processes with applications to biology}- 2nd Ed.(Chapman \& Hall-CRC, New York 2010).
\bibitem{Grasman}
J. Grasman and O. A. van Herwaarden, {\em Asymptotic methods for the Fokker-Planck equation and the exit problem in applications} (Springer, Berlin 1999).

\bibitem{Novo2022C}
A. E. Bernardini and O. Bertolami, {\em Distorted stability pattern and chaotic features for quantized prey-predator-like dynamics}, Phys. Rev. E {\bf 107}, 044201 (2023).

\bibitem{LV1}
A. J. Lotka, {\em Elements of physical biology} (Williams \& Wilkins Co., Baltimore 1925).
\bibitem{LV2}
V. Volterra, {\em Variazioni e fluttuazioni del numero d'individui in specie animali conviventi}, Mem. R. Accad. Naz. Lincei. (Ser. VI) 2, 31-113 (1926).
\bibitem{Este}
A. Grassi and M. Marin\~o, {\em A Solvable Deformation of Quantum Mechanics}, Sigma {\bf 15}, 025 (2019).
\bibitem{90}
V. Pasquier and M. Gaudin, {\em The periodic Toda chain and a matrix generalization of the Bessel function recursion relations}, J. Phys. A: Math. Gen. {\bf 25}, 5243 (1992).
\bibitem{89}
N. A. Nekrasov and S. L. Shatashvili, {\em Quantization of integrable systems and four dimensional gauge theories}, in XVI$^{th}$ International Congress on Mathematical Physics, 265-289 (World Scientific, Singapore 2010).
\bibitem{95}
N. Seiberg and E. Witten, {\em Electric-magnetic duality, monopole condensation, and confinement in N=2 supersymmetric Yang-Mills theory}, Nuclear Phys. B {\bf 426}, 19 (1994); {\em Erratum}, Nuclear Phys. B {\bf 430}, 485 (1994).
\bibitem{46}
A. Gorsky, I. M. Krichever, A. Marshakov, A. Mironov and A. Morozov, {\em Integrability and Seiberg-Witten exact solution}, Phys. Lett. B {\bf 355} 466 (1995).
\bibitem{80}
E. J. Martinec and N. P. Warner, {\em Integrable systems and supersymmetric gauge theory}, Nuclear Phys. B {\bf 459}, 97 (1996).
\bibitem{21}
S. Codesido, A. Grassi and M. Mari\~{n}o, {\em Spectral theory and mirror curves of higher genus}, Ann. Henri Poincar\'e {\bf 18}, 559 (2017).
\bibitem{49}
A. Grassi, Y. Hatsuda and Mari\~{n}o, {\em Topological Strings from Quantum Mechanics}, Ann. Henri Poincar\'e {\bf 17}, 3177 (2016).
\bibitem{Meu2018}
A. E. Bernardini, {\em Testing non-classicality with exact Wigner currents for an anharmonic quantum system}, Phys. Rev. A {\bf 98}, 052128 (2018).

\bibitem{Bio17}
R. Real, A. M\'arcia Barbosa and J. W. Bull, {\em Species Distributions, Quantum Theory, and the Enhancement of Biodiversity Measures}, Syst. Biol. {\bf 66}, 453 (2017).
\bibitem{Coffey07}
W. T. Coffey, Yu. P. Kalmykov, S. V. Titovac and B. P. Mulligana, {\em The Langevin Equation: With Applications to Stochastic Problems in Physics, Chemistry and Electrical Engineering}, Phys. Chem. Chem. Phys. {\bf 9}, 3361 (2007).
\bibitem{Gradshteyn}
I. S. Gradshteyn and I. Ryzhik, {\it Tables of Integrals, Series and Products} (Academic Press, New York 1994).

\bibitem{Scripta}
A. E. Bernardini, {\em Geometrical structure of the Wigner flow information quantifiers and hyperbolic stability in the phase-space framework}, Phys. Scripta {\bf 100}, 125230 (2025).
\bibitem{PRE-LV4}
T. Reichenbach, M. Mobilia and E. Frey, {\em Coexistence versus extinction in the stochastic cyclic Lotka-Volterra model}, Phys. Rev. E {\bf 74}, 051907 (2006).
\bibitem{Mac63}
M. L. Rosenzweig and R. H. MacArthur, {\em Graphical Representation and Stability Conditions of Predator-Prey Interactions}, Am. Nat. {\bf 97}, 209 (1963).
\bibitem{Valenti2004}
A. Fiasconaro, D. Valenti, B. Spagnolo, {\em Nonmonotonic Behaviour of Spatiotemporal Pattern Formation in a Noisy Lotka-Volterra System}, Acta Phys. Pol. B {\bf 35}, 1491-1500 (2004).
 
\bibitem{Valenti2016}
D. Valenti, G. Denaro, B. Spagnolo, S. Mazzola, G. Basilone, F. Conversano, C. Brunet, and A. Bonanno, {\em Stochastic models for phytoplankton dynamics in Mediterranean Sea}, Ecol. Complexity {\bf 27}, 84-103 (2016).

\bibitem{26}
A. Golubev, {\em Applications and implications of the exponentially modified gamma distribution as a model for time variabilities related to cell proliferation and gene expression}, J. Theo. Biology {\bf 393}, 203 (2016).
\bibitem{27}
A. Poon, B. H. Davis and L. Chao, {\em Coupon Collector and the Suppressor Mutation}, Genetics {\bf 170}, 1323 (2005).

\bibitem{Bord13}
M. Bordonaro and V. Ogryzko, {\em Nontrivial quantum and quantum-like effects in biosystems}, Biosystems {\bf 112}, 11 (2013).

\bibitem{Bord19}
M. Bordonaro, {\em Quantum biology and human carcinogenesis}, Biosystems {\bf 178}, 16 (2019).

\bibitem{DNA1}
M. Winokan, L. Slocombe, J. Al-Khalili and M. Sacchi, {\em Multiscale simulations reveal the role of PcrA helicase in protecting against spontaneous point mutations in DNA}, Scientific Reports {\bf 13}, 21749 (2023).

\bibitem{DNA2}
L. Slocombe, M. Sacchi and J. Al-Khalili, {\em An Open Quantum Systems approach to proton tunnelling in DNA}, Communications Physics {\bf 5}, 109 (2022).

\bibitem{Caldeira}
A. O. Caldeira and A. J. Leggett, {\em Path integral approach to quantum Brownian motion}, Phys. A: Stat. Mech. Appl. {\bf 121}, 587 (1983).

\bibitem{Parisi2023}
G. Parisi, {\em Nobel Lecture: Multiple equilibria}, Rev.\ Mod.\ Phys.\ {\bf 95}, 030501 (2023). 

\bibitem{Stassi2015}
R. Stassi, S. De Liberato, L. Garziano, B. Spagnolo, and S. Savasta,
{\em Quantum control and long-range quantum correlations in dynamical Casimir arrays}, Phys.\ Rev.\ A {\bf 92}, 013830 (2015).

\bibitem{Carollo2018}
A. Carollo, B. Spagnolo, and D. Valenti,
{\em Symmetric logarithmic derivative of fermionic Gaussian states}, Entropy {\bf 20}, 485 (2018).

\bibitem{Magazzu2015}
L. Magazz\'u, D. Valenti, A. Carollo, and B. Spagnolo,
{\em Multi-state quantum dissipative dynamics in sub-Ohmic environment: the strong coupling regime}, Entropy {\bf 17}, 2341-2354 (2015).

\bibitem{Spezia2008}
S. Spezia, L. Curcio, A. Fiasconaro, N. Pizzolato, D. Valenti, B. Spagnolo, P.
Lo Bue, E. Peri, and S. Colazza, {\em Evidence of stochastic resonance in the mating behavior of Nezara viridula (L.)}, Eur.\ Phys.\ J.\ B {\bf 65}, 453-458 (2008).

\bibitem{Agudov2021}
N. V. Agudov, A. V. Dubkov, A. V. Safonov, D. V. Guseinov, and M. Matyushkin,
{\em Stochastic model of memristor based on the length of conductive region}, Chaos Solitons \& Fractals {\bf 150}, 111131 (2021).

\bibitem{JSTAT2009}
D. Benedetti,
{\em Critical behavior in spherical and hyperbolic spaces}, J.\ Stat.\ Mech.\ P01002 (2009).

\bibitem{DiFresco2022}
G. Di Fresco, B. Spagnolo, D. Valenti, and A. Carollo,
{\em Multiparameter quantum critical metrology}, SciPost Phys.\ {\bf 13}, 077 (2022).

\end{thebibliography}
\end{document}